%% file: portrayal.tex
\documentclass[sigconf]{acmart}

\usepackage{multirow}
\usepackage{dirtytalk}
\usepackage{colortbl}
\usepackage{xcolor}
\copyrightyear{2023} 
\acmYear{2023} 
\setcopyright{rightsretained} 
\acmConference[DIS '23]{Designing Interactive Systems Conference}{July 10--14, 2023}{Pittsburgh, PA, USA}
\acmBooktitle{Designing Interactive Systems Conference (DIS '23), July 10--14, 2023, Pittsburgh, PA, USA}
\acmDOI{10.1145/3563657.3596000}
\acmISBN{978-1-4503-9893-0/23/07}

\newcommand{\toolname}{Portrayal}

\newcommand{\revise}{\textcolor{black}}

\begin{document}



\title[\toolname]{\textsc{\toolname}: Leveraging NLP and Visualization for Analyzing Fictional Characters}

\author{Md Naimul Hoque}
\email{nhoque@umd.edu}
\affiliation{
    \institution{University of Maryland, College Park}
    \city{College Park}
    \state{MD}
    \country{USA}
}

\author{Bhavya Ghai}
\authornote{Bhavya Ghai is with Amazon now.  He contributed to this work while he was with Stony Brook University.}
\email{bghai@cs.stonybrook.edu}
\affiliation{
    \institution{Stony Brook University}
    \city{Stony Brook}
    \state{NY}
    \country{USA}
}

\author{Kari Kraus}
\email{kkraus@umd.edu}
\affiliation{
    \institution{University of Maryland, College Park}
    \city{College Park}
    \state{MD}
    \country{USA}
}

\author{Niklas Elmqvist}
\email{elm@umd.edu}
\affiliation{
    \institution{University of Maryland, College Park}
    \city{College Park}
    \state{MD}
    \country{USA}
}




\renewcommand{\shortauthors}{Hoque et al.}

\begin{abstract}
    Many creative writing tasks (e.g., fiction writing) require authors to write complex narrative components (e.g., characterization, events, dialogue) over the course of a long story.
    Similarly, literary scholars need to manually annotate and interpret texts to understand such abstract components.
    In this paper, we explore how Natural Language Processing (NLP) and interactive visualization can help writers and scholars in such scenarios.
    To this end, we present \textsc{Portrayal}, an interactive visualization system for analyzing characters in a story. Portrayal extracts natural language indicators from a text to capture the characterization process and then visualizes the indicators in an interactive interface.
    We evaluated the system with 12 creative writers and scholars in a one-week-long qualitative study.
    Our findings suggest Portrayal helped writers revise their drafts and create dynamic characters and scenes.
    It helped scholars analyze characters without the need for any manual annotation, and design literary arguments with concrete evidence.
\end{abstract}



\begin{teaserfigure}
    \includegraphics[width=0.94\textwidth]{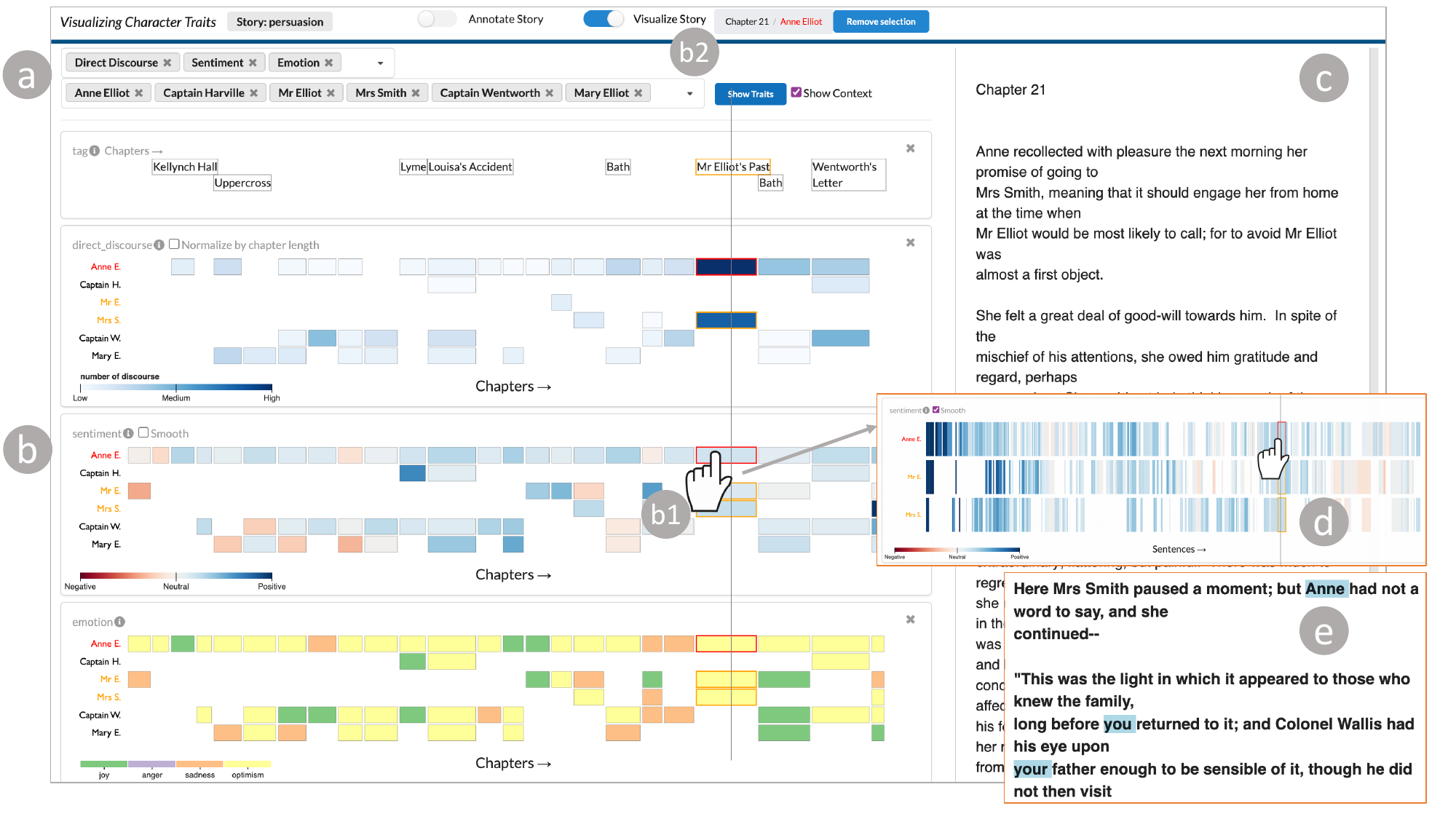}
    \caption{\textsc{Overview of \toolname{} using the novel \textit{Persuasion} by Jane Austen.}
    (a) A control panel for selecting characters and their trait indicators (e.g., actions).
    (b) The central panel shows the trait indicators for each character across the chapters.
    In this example, we see three indicators (direct speech or discourse, sentiment, and emotion) in three separate views, aligned vertically.
    Each row in these views represents a character while each column represents a chapter.
    The views are synchronized.
    For example, on hovering over a cell (b1), the system highlights the character of interest in red (Anne Elliot) and other co-occurring characters of that chapter in orange.
    The selected character and chapter (chapter 21 in this case) also appear in the top (b2). The text editor moves to the corresponding text editor (c).
    When a user clicks on a cell (b1), the views update to a sentence wise visualization (d) where each cell represents a sentence of the selected chapter.
    On hovering over a tile in this view (d), the system highlights the selected sentence and the mentions of the character in the text editor (e).}
    \Description{A figure showing a screenshot of Portrayal. Three rectangular boxes are placed vertically in the middle of the screen. The first box shows the number of direct speeches for the characters in a story with a collection of tiles. The leftmost tiles are from the first chapter, while the rightmost ones are from the last chapter. The tiles are color coded; the brighter the color, the more is the number of direct speeches. The second and third boxes have similar visual style except they represent sentiment and emotions of the characters. There is a text editor on the side of these three boxes. There are two sub-figures on the top of the text editor. The first one has colored tiles to show sentiment scores in a chapter. The second sub-figure shows a sentence.}
    \label{fig:teaser}
\end{teaserfigure}

\maketitle

\input{sections/1introduction}
\input{sections/2background}
\input{sections/3related_works}
\input{sections/4formative_study}
\input{sections/5design_guidelines}

\input{sections/6system}
\input{sections/7evaluation}

\input{sections/8discussion}
\input{sections/9conclusion}

\begin{acks}
    We thank the creative writers and scholars who participated in our evaluation. We also thank the anonymous reviewers for their thoughtful feedback.
\end{acks}

\bibliographystyle{ACM-Reference-Format}
\bibliography{portrayal}

\end{document}

%% file: sections/1introduction.tex
\section{Introduction}

\revise{Writing and understanding literature are two fundamentally human activities, but today we have the potential to use natural language processing (NLP) methods to help in both endeavors.}
For the writer, recent tools go beyond mere spelling and grammatical support to the use of Large Language Models (LLMs) to support human-AI co-writing~\cite{DBLP:conf/ACMdis/GeroLC22, talebrush, lee2022coauthor, yuan2022wordcraft}.
In this paradigm, a writer asks the AI to generate a section of the story based on a prompt; the writer then edits the generated story and provides further prompts.
Scholars and critics, on the other hand, use language to understand society and the literature borne from it.
For example, they use Voyant Tools~\cite{voyant} and Google Ngram Viewer~\cite{lieberman2007quantifying} to automatically find the most frequent words and phrases in a corpus.
While existing tools are useful to writers and scholars, it is still unknown how to best apply NLP to more complex and abstract literary and narrative components such as characterization, dialogue, and narrative structures, and how that can help writers and scholars.

To explore this design space, we present a study of how characters, one of the central components to narrative fiction~\cite{rimmon2003narrative}, can be extracted from a text and represented visually in an interactive environment.
Many writers base their stories on rich and complex characters that drive the plot, almost as if they have a life of their own.
As a case in point, witness Walter White slowly transforming from milquetoast chemistry teacher to insidious drug kingpin in AMC's \textit{Breaking Bad} (2008); Edmond Dant{\`e}s evolution from a na{\"i}ve and an innocent young man to a calculating and vengeful nobleman in \textit{The Count of Monte Cristo} by Alexandre Dumas (1844); or Samwise Gamgee metamorphizing from Frodo Baggins's gardener and best friend into his stalwart champion and protector in J.R.R.\ Tolkien's \textit{The Lord of the Rings} (1954).
Designing engaging and dynamic characters is challenging for writers as they need to believably show the evolution of the character throughout the story.
While the process for developing characters varies from writer to writer, we believe an analytic tool at the time of editing or revising a draft can help writers to identify discrepancies between intended and written text, deciding what edits to make, and how to make those desired edits.
Such analytic support could also help literary scholars, who manually annotate the text (i.e., close reading~\cite{boyles2012closing}) for understanding characters or for writing a character analysis essay.

We propose \textbf{\textsc{\toolname}}, a web-based interactive system for analyzing characters in a story.
\revise{The overarching goal of the tool is to surface character patterns and properties from a written text.
We believe such a tool has the potential to support diverse creative writing and analysis tasks.
The two most relevant user groups and tasks for such a tool are (1) writers, who want to create dynamic and engaging characters, and (2) scholars, who want to analyze, interpret, and critique characters.}
To inform our research, we reviewed existing literature on narrative fiction, NLP, and visual analytics (\S\ref{sec:background} and \S\ref{sec:related_work}) and conducted a formative study (\S\ref{sec:formative}) with three creative writers, three literary scholars, and two participants who were both creative writers and literary scholars. 
The key features of \toolname{} include (1) a text editor where writers can upload their partial or full draft while scholars can upload a text written by others (Figure~\ref{fig:teaser}c); (2) an analytic pipeline that can extract various trait indicators (e.g., actions, emotions, speech, and external appearance) of the characters that capture their journey over the course of the story (a process referred to as \textit{characterization}~\cite{rimmon2003narrative}); and (3) an interactive visual interface for visualizing the trait indicators, where users can investigate single or multiple indicators together (Figure~\ref{fig:teaser}).

To evaluate \toolname{}, we conducted a user study involving four creative writers, four literary scholars, and four participants who were both.
Writers in our study used \toolname{} to analyze one of their unpublished drafts independently for a week.
In a debriefing meeting at the end of the week, we asked them to describe any new insights or thoughts arising from using \toolname{}.
Literary scholars followed a similar protocol, except they were asked to analyze two stories among four well-known and representative stories.
Our findings suggest that \toolname{} helped writers (1) create dynamic characters that go through many emotional changes, (2) create dynamic scenes where characters with opposite sentiments and emotions interact, (3) evaluate character arcs, and (4) find unintentional prioritizing of a group of characters with similar social identity (e.g., female characters).
Literary scholars reported that \toolname{} helped them find tangible evidence to support their literary arguments as well as find the lack of evidence for an opposing argument.
Further, the scholars noted that \toolname{} could help young scholars write literary essays in coursework and would work as a probe to facilitate literary conversation and debate.
In summary, our contributions are as follows:

\begin{enumerate}

    \item A formative study with eight creative writers and literary scholars to identify requirements for understanding and developing characters using NLP and visualization;
    
    \item The design and development of \textsc{\toolname{}}, an interactive system to automatically detect and visualize natural language indicators of character traits; and
    
    \item A user study with 12 creative writers and literary scholars to report the application and potential of \toolname{}.
    
\end{enumerate}

%% file: sections/2background.tex
\section{Background: Narrative Fiction and Characters}
\label{sec:background}
A \textit{narrative fiction} is ``the narration of a succession of fictional events''~\cite{rimmon2003narrative}. Such narratives are composed of several components (e.g., events, characters,
time, causality, focalization, and narration style). This paper focuses on characters, an essential component of narrative
fiction.
According to Chatman~\cite{chatman1980story}, a \textit{character} is a construct within an abstracted story that is described through a network of personality traits (e.g., Sarrasine is feminine, Othello is jealous, Roland Deschain is brave).
However, these traits may or may not be present in the text directly; they are often displayed and exemplified through various \textit{indicators}.
The process of describing the character traits by designing the indicators in the text is called \textit{characterization}.

Rimmon-Kenan~\cite{rimmon2003narrative} proposed two types of indicators to understand characterization: 
(1) \textit{direct definition} are indicators that explicitly mention traits (e.g., ``John is \textit{stressed}.''); and (2) \textit{indirect definition} are those that do not mention the traits explicitly but rather indirectly refer to them through activity and examples (e.g., ``John tapped his foot and looked at his watch.'').
The latter, characterization through what the character says and does rather than the writer directly stating these traits, is known as ``show, don't tell'' and is an almost universal rule of thumb in fiction writing.
Rimmon-Kenan~\cite{rimmon2003narrative} proposed four main methods for indirect definition: actions, speech, appearance, and environment.

\paragraph{Actions.}

Habitual actions or one-time actions can imply a trait.
One-time actions tend to evoke dynamic aspects of a character, often playing a pivotal role in a narrative.
In contrast, habitual actions tend to evoke the static aspects of a character.
For example, Gandalf's habitual actions in Tolkien's \textit{The Lord of the Rings} (1954) include his occasional pipe smoking and wizardly brooding.
Sam Spade has a gruff and cynical demeanor\footnote{Hammett also calls him ``a blonde satan,'' which is perhaps a little more on the nose than what is common.} in Dashiell Hammett's \textit{Maltese Falcon} (1930).
As for one-time actions, consider the following significant passage from Chapter 2 in \textit{Hunger Games} (2008) by Suzanne Collins that cements our view of Katniss Everdeen as both brave and loving of her sister Prim: 

\begin{quote}
    \tt
    ``Prim!''
    The strangled cry comes out of my throat, and my muscles
    begin to move again.
    ``Prim!''
    I don't need to shove through the crowd.
    The other kids make way immediately allowing me a straight path to the stage.
    I reach her just as she is about to mount the steps. With one sweep of my arm, I push her behind me.
    ``I volunteer!'' I gasp. ``I volunteer as tribute!''
\end{quote}

\paragraph{Speech.}

Speech is an essential indicator for characterization.
Consider the following direct speech by the lead character Anne Elliot in Jane Austen's \textit{Persuasion} (1817), which reveals her confidence, maturity, and determination to make her own decisions that have only come with age:

\begin{quote}
    \tt
    ``You should not have suspected me now; the case is so different, and my age is so different.
    If I was wrong in yielding to persuasion once, remember that it was to persuasion exerted on the side of safety, not of risk.
    When I yielded, I thought it was to duty, but no duty could be called in aid here.
    In marrying a man indifferent to me, all risk would have been incurred, and all duty violated.''
\end{quote}

\paragraph{External Appearance.}

External appearances such as eye color, hairstyle, and clothing can indicate a character's personality.
Consider the description of Mr.\ Wednesday from Neil Gaiman's \textit{American Gods} (2001):

\begin{quote}
    \tt
    His hair was reddish-grey; his beard, little more than stubble, was greyish-red.
    A craggy, square face with pale grey eyes. 
    The suit looked expensive, and was the colour of melted vanilla ice-cream.
    [...]
    There was something strange with his eyes, Shadow thought.
    One of them was darker than the other.
\end{quote}

\paragraph{Environment.}

A character's surrounding physical environment (e.g., room, house, street) and the human environment (family, colleagues, and social class) can indicate traits.
Consider the following description of Mr.\ Phileas Fogg, Esq.'s daily haunts from Jules Verne's \textit{Around the World in Eighty Days} (1872): 

\begin{quote}
    \tt
    He lived alone in his house in Saville Row, whither none penetrated.
    A single domestic sufficed to serve him.
    He breakfasted and dined at the club, at hours mathematically fixed, in the same room, at the same table, never taking his meals with other members, much less bringing a guest with him; and went home at exactly midnight, only to retire at once to bed.
    [...]
\end{quote}

Note that there is another category of indicator, \textit{analogy}, which does not introduce any new indicator but instead reinforces a previously defined trait by analogy.
An example would be stating that a character ``is just like his brother.''

Our goal in this paper is to study how these indicators can be extracted using NLP and visualized in an interactive environment.

%% file: sections/3related_works.tex
\section{Related Work}
\label{sec:related_work}

In this section, we provide an overview of analytic and visualization tools for creative writing and literary analysis.

\subsection{Computational Support for Creative Writing}

Modern creative writers often use a range of computational tools.
The first kind is text editor-like writing software such as Microsoft Word or Google Docs. 
There are several professional and paid software available to writers.
For example, Scrivener~\cite{scrivener} allows writers to organize a story in sections, add synopsis and notes to each section, and easily merge or swap between sections.
Granthika~\cite{granthika} is a similar sort of paid service where writers can outline a more detailed narrative world, including character descriptions, major events in a timeline, and causal constraints on the events.
These tools primarily enhance organizational capabilities of writers with limited feedback on the actual writing.

More recently, researchers have proposed several emerging writing support tools.
A dominant trend is the use of LLMs. 
These tools can generate text based on a prompt, often helping writers explore alternate narrative worlds and creative angles~\cite{brown2020language, talebrush, lee2022coauthor, yuan2022wordcraft, DBLP:conf/ACMdis/GeroLC22, mirowski2022cowriting}.
However, many open issues remain for LLMs based co-writing paradigm such as the lack of capability to generate long coherent texts, generated text often being perceived as formulaic by writers, and lack of trust in the machine~\cite{mirowski2022cowriting}.
Another set of tools use NLP to extract patterns from the text that are otherwise difficult to notice.
These tools typically focus on providing feedback while writers revise their text.
For example, Du et al.~\cite{du-etal-2022-read, du-etal-2022-understanding-iterative} proposed a tool to help writers in their iterative revising process.
Sterman et al.~\cite{sterman2020interacting} proposed an analytic model that allows writers to interact with the literary style of an article.
DramatVis Personae~\cite{DBLP:conf/ACMdis/HoqueGE22} helps writers mitigate nuanced social biases in their writing.
Dang et al.~\cite{DBLP:conf/uist/DangBLB22} integrated automatic text summarization in a text editor to help writers revise analytical essays.

While existing tools are useful, we are not aware of any tools that provide analytic feedback to writers for developing characters, or any other narrative components. 
Writing fiction is a complex task, often requiring writers to manage several narrative components such as characters, places, and events to design a engaging story.
We believe analytic support for these narrative components could help writers.
To demonstrate our research proposal, we propose an interactive system to represent \textit{characters}, one of the central component to fiction.
Our work shows that such a system could help writers in their iterative writing process, identifying patterns and idiosyncrasies that they were unaware of during writing. 
Another unique feature of our tool is the use of interactive visualization, which enabled us to represent characters visually and help writers use analytic support without the need for algorithmic expertise.

\subsection{Computational Support for Literary Analysis and Digital Humanities}
\label{sec:viz_humanities}

Literary analysis can be divided into two broad areas: \textit{close} and \textit{distant} reading.
Close reading~\cite{boyles2012closing} is the traditional method for analyzing a text.
In this method, scholars carefully read and annotate individual words and sentences (e.g., coloring, underlying words) to unravel syntax, semantics, and formal structure.
It is an argumentative process where scholars express their interpretation of a text with excerpts from the text supporting their arguments.
Distant reading, on the other hand, focuses on finding patterns and statistics from a large corpus of texts without the need to extensively read the text.
Scholars typically rely on programming and text analytics tools for distant reading.
Voyant Tools~\cite{voyant}, Google Ngram Viewer~\cite{lieberman2007quantifying}, Hedonometer~\cite{reagan2016emotional}, and  Wordle~\cite{DBLP:journals/tvcg/ViegasWF09} are examples of tools that support distant reading. 

While distant reading helps scaling up analysis to hundreds of texts, scholars still rely on close reading for critically analyzing characters.
To do so, scholars need to manually identify the character traits and patterns to write their interpretation (see this blog post~\footnote{\url{https://www.wikihow.com/Write-a-Character-Analysis}} for a detail overview) of the characters.
There are limited computational support for this task. One notable tool is eMargin~\cite{kehoe2013emargin}, providing an interface where scholars can collaboratively annotate a text. However, it does not provide any support to understand character traits or other narrative components.
Our work aims to help scholars analyze nuanced character traits without the need of extensive annotation of the text, a novel approach in the digital humanities domain.

\begin{figure*}[t]
    \centering
    \includegraphics[width=0.9\linewidth]{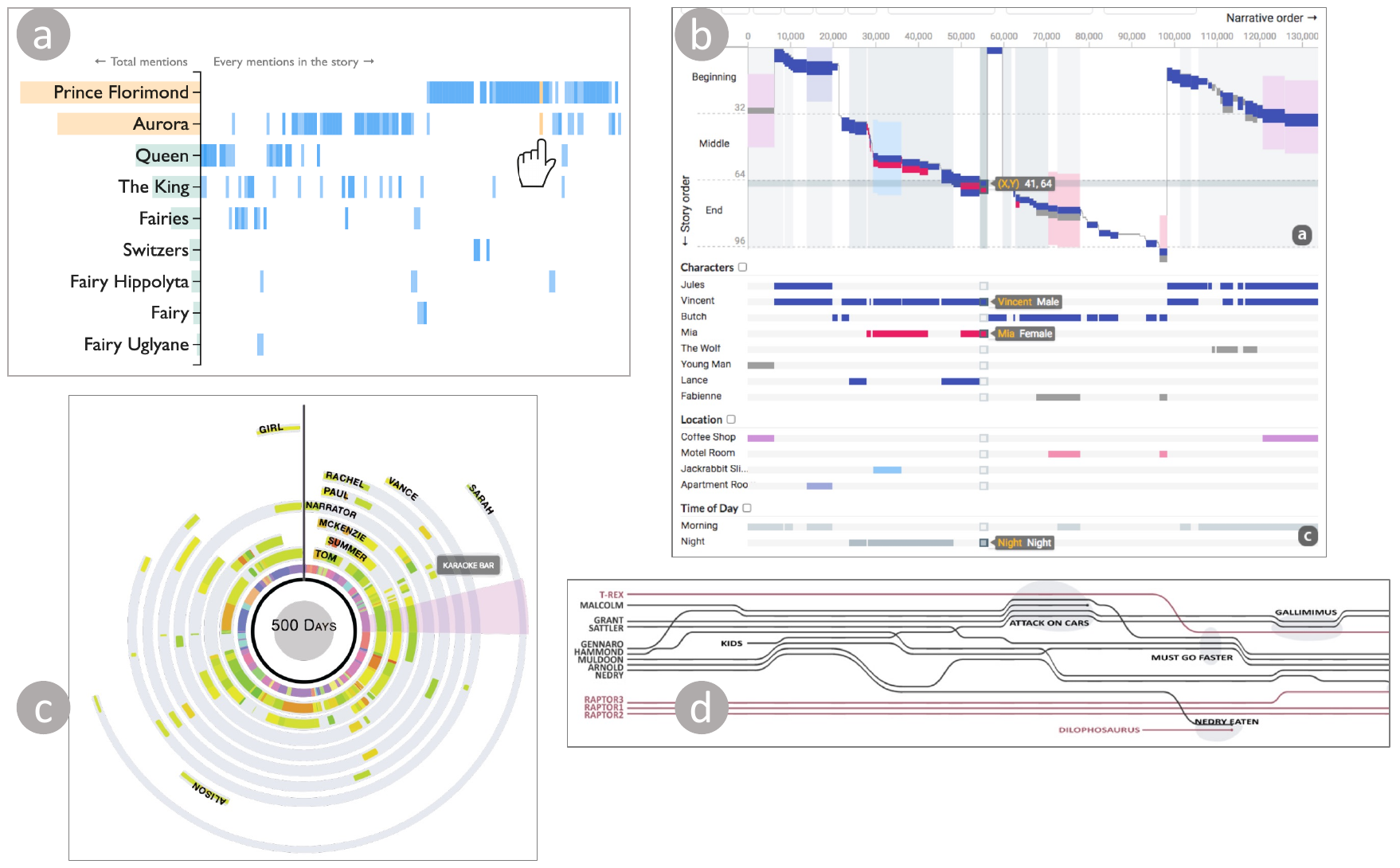}
    \caption{\textsc{Visualizations for narrative analysis (formative study).}
    (a) DramatVis Personae (dvp)~\cite{DBLP:conf/ACMdis/HoqueGE22}; (b) StoryCurves~\cite{DBLP:journals/tvcg/KimBISGP18}; (c) StoryPrint~\cite{DBLP:conf/iui/WatsonSSGMK19}; and (d) StoryLine~\cite{DBLP:journals/tvcg/LiuWWLL13, DBLP:journals/tvcg/TanahashiM12, DBLP:conf/gd/GronemannJLM16}.
    The images are used with permission from the respective authors.}
    \Description{A figure showing screenshots for four existing narrative visualizations.}
    \label{fig:formative_pictures}
\end{figure*}

\subsection{Visualizing Text and Narrative Components}
\label{sec:design}

Visualization can reveal hidden patterns in a text~\cite{alencar2012seeing}.
This has fueled techniques to visualize and summarize a document or corpus.
Word clouds, a truly community-driven (``vernacular'') form of text visualization~\cite{DBLP:journals/tvcg/ViegasWF09}, are one of the most common text summarization technique and appear regularly on the web.
Due to the wide applications of word clouds, researchers have proposed methods to improve them~\cite{hearst2019evaluation, DBLP:journals/cga/CuiWLWZQ10, DBLP:journals/tvcg/LeeRKC10}.
Extending this line of work, TextFlow~\cite{DBLP:journals/tvcg/CuiLTSSGQT11} and ThemeDelta~\cite{DBLP:journals/tvcg/GadJGEEHR15} proposed two novel interactive visualizations for understanding evolving topics in a text corpus. 
FacetAtlas~\cite{DBLP:journals/tvcg/CaoSLGLQ10} visualizes relation between different facets of a corpus. Text visualization can also help specific applications.
For example, Jigsaw~\cite{DBLP:conf/ieeevast/StaskoGLS07} helps investigative journalists link various entities and documents. Poemage~\cite{DBLP:journals/tvcg/McCurdyLCM16} helps literary scholars understand sonic properties in a poem.
For a detailed overview of text visualization applications for digital humanities, we refer readers to the survey conducted by J{\"{a}}nicke et al.~\cite{DBLP:conf/vissym/JanickeFCS15}.

In our context, there are four visualizations that are most relevant to our work (Figure~\ref{fig:formative_pictures}).
StoryLine~\cite{DBLP:journals/tvcg/LiuWWLL13, DBLP:journals/tvcg/TanahashiM12, DBLP:conf/gd/GronemannJLM16} focuses on grouping characters in a timeline to highlight plotlines in a narrative.
StoryCurves~\cite{DBLP:journals/tvcg/KimBISGP18} helps scholars analyze non-linear narrative structures, while StoryPrint~\cite{DBLP:conf/iui/WatsonSSGMK19} helps filmmakers organize movie scenes and the characters in it.
DramatVis Personae (DVP)~\cite{DBLP:conf/ACMdis/HoqueGE22} uses a timeline and a word zone to help writers identify social biases. 
While each of these works uses characters to represent texts visually, none of them help literary scholars and writers to analyze characters, their traits or characterization.
We study how to conceptualize characters using a computational model and visualize them in a flexible and interactive tool, a novel approach for creativity support tool design. 

%% file: sections/4formative_study.tex
\begin{table*}[t]
    \centering
    \begin{tabular}{lllp{2.2cm}p{6.8cm}c}
    \toprule
    \rowcolor{gray!10} 
    \textsc{\textbf{Id}} & \textsc{\textbf{Gender}} & \textsc{\textbf{Age}} & \textsc{\textbf{Profession}} & \textsc{\textbf{Experience}} & \textsc{\textbf{Yrs Exp}} \\ 
    \midrule
    
    P1 & Female & 18-24 & Literary scholar & BA and MA in English Literature & 7  \\
    \rowcolor{gray!10} 
    P2 & Female & 25-34 & Literary scholar & Doctoral student focusing at European and African Literature (Anglophone and Francophone) & 10 \\
    
    P3 & Male & 25-34 & Literary scholar & BA and MA in English Literature; Instructor & 12 \\
    
    \midrule
    \rowcolor{gray!10} 
    
    P4 & Female & 25-34 & Creative writer & Novels (fiction/non-fiction), short stories, screenplays, poems, blogs, critiques, and fanfiction & 15 \\
    
    P5 & Female & 45-54 & Creative writer & Novels (fiction/non-fiction) & 25 \\
    
    \rowcolor{gray!10} 
    P6 & Male & 25-34 & Creative writer & Short stories, poems, and fiction & 12 \\
    
    \midrule
    
    P7 & Male & 35-44 & Literary scholar \& Creative writer & Doctoral student with interest in using text mining for contemporary Persian literature; published novelist & 18 \\
    
    \rowcolor{gray!10}
    P8 & Male & 45-54 & Literary scholar \& creative writer & MA in English and MFA in creative writing; Instructor at a writing institute; Short stories, fiction & 20 \\

    
    \bottomrule
    \end{tabular}
    \caption{\textsc{Participant demographics for the formative study.}
    The participants can be divided into three main groups: literary scholars (top), creative writers (middle), and participants who are both (bottom).
    }
    \Description{A table showing demographic information of eight participants. The table has six columns: ID, Gender, Age, Profession, Experience, and Years of Experience.}
    \label{tab:participant_formative}
\end{table*}
\section{Formative Study}
\label{sec:formative}

We conducted semi-structured interviews with creative writers and scholars to understand their current workflow, process for developing or analyzing characters, and potential for computational support in this regard.
Our university's Institutional Review Board (IRB) approved the study.

\subsection{Participants}

We recruited three literary scholars, three creative writers, and two participants who were both literary scholars and creative writers (Table~\ref{tab:participant_formative}).
We recruited participants by advertising in our university's Department of English, literary center, and on other relevant campus mailing lists.
Our inclusion criteria for literary scholars included academic training in literature, such as having a bachelor's or post-graduate degree in English literature and creative writing and/or having research experience with literary analysis. 
For creative writers, we required participants to have published stories in their portfolios.
Note that literary scholarship and creative writing are not mutually exclusive---a person can be both.
Thus, we have three sets of participants---literary scholars (P1-P3), creative writers (P4-P6), and participants who were both (P7-P8). 

\subsection{Procedure}

We conducted the interviews over Zoom.
Each interview lasted around 1 hour and was divided into three parts.
First, after gathering informed consent and a brief introductory discussion, we asked participants to share their current workflow for writing or analyzing. In the second part, we asked writers how they develop characters while we asked scholars to share how they analyze characters.
Finally, we showed participants a few screenshots from current narrative visualization solutions (Figure~\ref{fig:formative_pictures}). All of them used characters as encoding in the representation, although none of them were specifically designed for characterization.
To demonstrate how interactivity is operationalized in the existing solutions, we showed a short online demo of DramatVis Personae~\cite{DBLP:conf/ACMdis/HoqueGE22}.
We asked participants their opinion about the visualizations, focusing on how similar sort of tools can help them develop or analyze characters.
This part was designed to direct participants toward our research focus and incite discussion about the design of our future visualization tool.
Participants brainstormed with the study administrator and provided suggestions. 
At the end of the session, we compensated each participant with a \$20 gift card. We provide the questionnaire for the interviews as a supplement.


\subsection{Analysis}

We created an anonymized transcript for each interview from the recorded audio.
Two authors of this paper open-coded the transcripts independently.
A code was generated by summarizing relevant phrases or sentences from the transcripts with a short descriptive text.
Both coders then conducted a thematic analysis~\cite{braun2006using} to group related codes into themes.
Coders met regularly to discuss disagreements and refine the codes and themes iteratively. 
The codes and themes were also regularly discussed with the entire research team.

\subsection{Findings}

Our findings relate to the topics of narrative fiction, characters, their traits, and visual representation.
We discuss the findings below.
We link the design requirements identified from the interviews wherever applicable. 






\begin{table*}[t]
    \centering
    \begin{tabular}{llp{7.8cm}ll}
    \toprule
    \rowcolor{gray!10} 
    \textbf{\textsc{Category}} & \textbf{\textsc{Id}} & \textbf{\textsc{Design Requirement}} & \textbf{\textsc{Origin}} & \textbf{\textsc{Participants}} \\ 
    \midrule
    
    \multirow{5}{2cm}{Trait requirements} & R1 & Show presence of the characters & Formative study & P1-8 \\
    
    & R2 & Support analysis of actions & Formative study \& theory & P2, P4-8 \\
    
    & R3 & Support speech analysis & Formative study \& theory & P1-3, P5 \\
    
    & R4 & Support analysis for direct definition and external appearance &  Formative study \& theory & P3, P5, P8\\
    
    & R5 & Support sentiment analysis & Formative study & P2-5, P7-8 \\
    
    & R6 & Show environments for the characters & Theory & --- \\
    
    \midrule
    
    \multirow{5}{2cm}{System and visualization requirements} & R7 & Support analysis of multiple traits & Formative study \& theory & P1-8 \\
    
    & R8 & Show change and transformation for the traits & Formative Study & P4-7 \\
    
    & R9 & Linking text and visualization & Formative Study & P1-8 \\

     & R10 & Provide context and annotation for the visualization  & Formative study & P2-3, P6\\
    
    & R11 & Use linear timeline to represent the narrative  & Formative study & P1-2, P6-7 \\
    
    \bottomrule
    \end{tabular}
    \caption{\textsc{Design requirements.}
    These requirements were identified from the formative study and characterization theory.}
    \Description{A table listing eleven design requirements. The table has five columns: Category, Id, Design Requirement, Origin, and Participants.}
    \label{tab:design_requirements}
\end{table*}

\subsubsection{Understanding Workflow}

We identified several common workflows for both writers and scholars.

\paragraph{Current Writing Workflow for Developing Characters.}

Writers reported that they predominantly develop stories by showing a journey for the characters or by building an imaginary world through the lens of the characters.
Writers  mentioned the importance of characterization in this process: often the writing starts with thinking about what will be the physical appearance of the characters (P4-7), what will be their emotional state (P5, P7-8), and how they will react to certain important events (P4-6).
However, most writers in our study did not report any specific workflow for characterization or developing the story for that matter.
This was expected given the creative nature of the task. 

\paragraph{Current Scholarly Workflow for Analyzing Characters.}

All scholars reported close reading as their method for analyzing a story and the characters within it.
Scholars mentioned that they often try to interpret the personality of characters (P1-2, P8), their actions (P1-3, P7), and language use (P3) from the text.
This analysis often results in an essay where they report their interpretation and provide portions of the text as evidence. Participants were  not aware of any tools for helping them in this process.
Additionally, scholars reported the annotation task (identifying important words and phrases) to be the most time-consuming and painstaking in close reading.

\subsubsection{How to Computationally Model Characters?}

During the brainstorming phase of the interviews, several participants were at first confused about how characters can be represented computationally and how that can help them. 
Screenshots of existing visualizations (Figure~\ref{fig:formative_pictures}) helped them understand our research proposal.
Writers speculated that such a tool for visualizing character traits could help them identify both intentional and unintentional characterizations.
They thought such a tool would be most useful during revising when reorganizing their thoughts and trying to find a better line of delivery.
Scholars noted that such a tool could work as a visual index and help them analyze character traits more thoroughly.
Participants further provided requirements and suggestions for such a tool, which are discussed next. 

\subsubsection{System Requirements}

This section outlines the system requirements identified from the interviews.

\paragraph{Support Analysis for Presence, Actions, Speech, Direct Definition, and Sentiment Analysis.}
Participants found character presence, which is common in all existing narrative visualizations, to be useful for understanding scenes and the overall structure of the story (P1-8) (\textbf{Requirement 1/R1}).
Moreover, most participants mentioned that actions are essential to designing and understanding characters (P2, P4-8).
Participants remarked that this important indicator is missing from the existing solutions (\textbf{R2}).
 P5 said:

\begin{quote}
    \texttt{``My characters tend to change their personality and actions throughout a story.
    I am not sure how you can show this, but if I can study the actions for my characters across a story, that will help me develop them.''} (P5)
\end{quote}

Matching the theory of characterization (Section~\ref{sec:background}), we found that speech (P1-3, P5) ( \textbf{R3}) and direct definition (P3, P5, P8) (e.g., adjectives and adverbs, \textbf{R4}) are important indicators. 
Finally, several participants found sentiment analysis to be useful in the existing solutions (StoryPrint and StoryCurves in Figure~\ref{fig:formative_pictures}b and~\ref{fig:formative_pictures}c) (P2-5, P7-8) (\textbf{R5}).
It is helpful to understand the emotional state of a character.

\paragraph{Analyze Multiple Traits Together.}

A recurring theme in our interview is that characters are multifaceted or multi-dimensional, and different trait indicators need to be studied together to understand a character (P1-8) (\textbf{R7}).
All participants found the existing visualizations helpful.
However, they identified two key limitations:
(1) the current list of indicators is insufficient for studying characters, and (2) current tools only allow the user to study one or at most two trait indicators together at any time.

\paragraph{Show Change and Transformation.}

Change or transformation of character traits is important for a narrative (P4-7) (\textbf{R8}).
Writers often try to depict a journey for the characters (i.e., character arc).
Participants thought sentiment analysis and actions were good candidates for showing the character arc.
P4 put it this way:

\begin{quote}
    \texttt{``We all know about the arc: a character starts as miserable and then transforms into something amazing.
    However, I try to avoid this traditional arc.
    My arcs are difficult to conceptualize and often hard to visualize in mind.
    In such a scenario, it will be interesting to see how the sentiment changes for a character in the full story.''} (P4)
 \end{quote}
 
\paragraph{Linking Text and Visualization.}

Participants suggested that to interpret the visualizations and integrate them into their work, our future tool should closely link the actual text and visualizations (\textbf{R9}). 
This will help writers validate feedback on their writing while scholars can use the links as annotations for close reading.

Additionally, participants suggested adding contextual information about the fiction in the visualizations themselves for improving interpretability (P1-2, P6-7) (\textbf{R10}). For example, P1 said:

\begin{quote}
    \texttt{I think summary texts over the visualization could make them more intuitive.} (P1)
\end{quote}

P7, who has both a literary and computational background, appreciated that StoryCurves (Figure~\ref{fig:formative_pictures}b) showed the location and time of the scenes.
However, they also mentioned that it might be difficult to extract such meaningful information from fiction.

\paragraph{Linear Timelines are Intuitive and Easy to Read.}

Participants found the linear, left-to-right timelines (StoryCurves, DramatVis Personae, and StoryLine), to be more intuitive than the radial timeline (Figure~\ref{fig:formative_pictures}c) (\textbf{R11}).
P2, P3, and P6 found similarities between linear timelines and dispersion plot~\cite{jacobs2018gutenberg}, a plot they use to understand where some search items appear in a corpus.
To add to that point, participants appreciated that StoryCurves (Figure~\ref{fig:formative_pictures}b) visualized non-linear narrative, which could be very helpful in writing or analyzing experimental narrative structure.
However, P7, who has a combined literature and computational background, suggested that contrary to a movie script (domain of StoryCurves) that has a pre-defined structure, extracting such a non-linear structure from fiction (our domain) would be difficult.

%% file: sections/5design_guidelines.tex
\section{Design Guidelines}

Based on our literature review (\S\ref{sec:background} and \S\ref{sec:related_work}) and the formative interviews (\S\ref{sec:formative}), we identified 11 design requirements for supporting the design and evaluation of characterization in a visual interface (Table~\ref{tab:design_requirements}).
In response, we identified the following design guidelines:

\begin{figure*}
    \centering
    \includegraphics[width=0.95\linewidth]{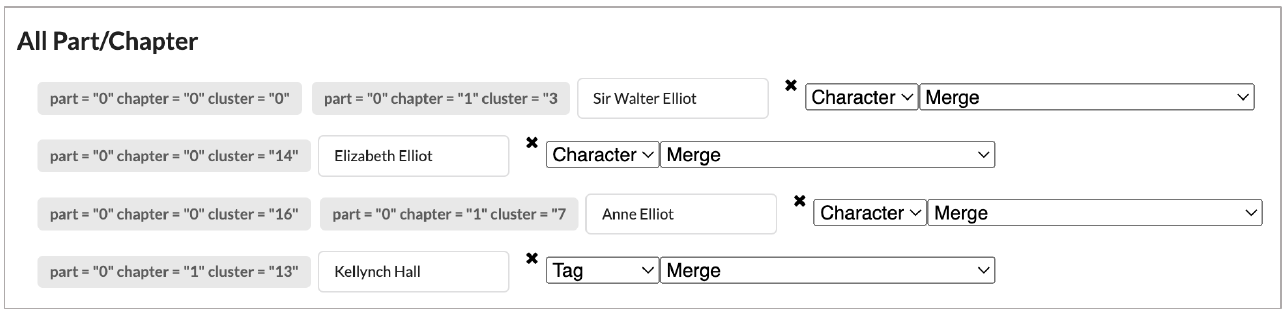}
    \caption{\textsc{Annotation interface.}
    This interface is used to annotate character mentions in a story.
    It allows assigning a name to a detected co-reference cluster, labeling the cluster as a character or contextual information, and merging a cluster with a previously detected cluster.
    For example, in the first row, two clusters from chapter 0 and 1 for the same character ``Sir Walter Elliott'' are merged.
    The clusters are connected to the text (not shown), allowing the user to validate the clusters.}
    \Description{A figure showing an interface of the tool. Each row in the interface indicates a co-reference cluster. Each cluster has several tags, a name, a dropdown indicating whether it's a character or not, and a dropdown to merge the cluster with another cluster. }
    \label{fig:annotate_interface}
\end{figure*}

\begin{itemize}
    \item[DG1] \textbf{Supporting Analysis for the Trait Indicators.}
    We identified six trait indicators that our system should support: presence, actions, speech, direct definition, sentiment analysis, and environments (\textbf{R1-R6}).
    Indicators such as presence (\textbf{R1}) and sentiment (\textbf{R5}) were found useful by participants from the previous visualization solutions, whereas \textbf{R2-R4} were identified by both participants and characterization theory.
    Finally, we did not find environments (\textbf{R6}) to be a major theme in our study.
    Nonetheless, it is listed as an indicator in characterization theory.
    Thus, we include it as a requirement. 

    \item[DG2] \textbf{Multiple Views Display.}
    To support \textbf{R7}, we need a design where each indicator can be studied in isolation as well as with other indicators.
    Additionally, we anticipated that the trait indicators we identified (\textbf{R1-R6}) might not be exhaustive.
    Thus, the design should be extensible to allow adding new indicators.
    We decided that the interface should follow a multiple view display~\cite{tufte1985visual}, where each indicator will have its own view, but they can be stacked together on top of each other for composite analysis.
    In the case of a newly discovered indicator, we can simply add a new view for that indicator.
    For supporting \textbf{R11}, the views should align vertically to have the same timeline.

    \item[DG3] \textbf{Show Transformation and Change for Indicators.}
    Change is important for a narrative.
    To support \textbf{R8}, our system should be able to extract and visualize the indicators and their changes over time.
    Based on the formative study, change is most relevant to actions (\textbf{R2}) and sentiment (\textbf{R5}).

    \item[DG4] \textbf{Linking Visualization and Text.}
    To support close reading (\textbf{R9}) and analysis of the indicators (\textbf{R1-R6}), the visualization should be closely linked to the text.
    We will extend interaction mechanism proposed in prior text visualization solutions~\cite{DBLP:conf/ACMdis/HoqueGE22, DBLP:journals/tvcg/KimBISGP18, DBLP:conf/ieeevast/StaskoGLS07} to support that. 

    \item[DG5]\textbf{Add Contextual Information.}
    We will add contextual information as an overlay to the visualization to enhance the semantics of the visualization (\textbf{R10}).
    We anticipate that contextual information can also convey environmental information (\textbf{R6}) about the characters.

\end{itemize}

We also note here a fundamental design guideline for the tool: it  \textbf{should not} try to replace human judgment in literary analysis and creative writing; rather, it should try to \textit{augment} human abilities in this endeavor.
Thus, the tool should be built on a human-in-the-loop architecture.

%% file: sections/6system.tex
\section{The \textsc{\toolname{}} System}
\label{sec:tool}

\toolname{} is a web-based visual analytics system for visualizing character traits in fiction.
Here we describe the two core components: (1) the NLP extraction pipeline; and (2) the visual interface.  

\subsection{Natural Language Processing Pipeline}

We designed the NLP pipeline to mainly extract the indicators from the text (\textbf{R1-R6}). The pipeline builds upon existing models and methods in NLP. We constructed a test dataset consisting of 10 sample books from Project Gutenberg to empirically evaluate the models and take design decisions. 

\begin{table}[t]
    \centering
    \begin{tabular}{p{1.5cm}p{6cm}}
    \toprule
    Sentence & \texttt{John laughs at me, of course, but one expects that in marriage.}\\
    \midrule
    \multirow{2}{*}{Proposition} & \textbf{\texttt{John} (ARG0)}; \textbf{\texttt{laughs} (V)}; \texttt{at me} (ARG2); \texttt{of course} (ARGM-DIS), but
     \textbf{one (ARG0)}; \textbf{\texttt{expects} (V)}; \texttt{that in marriage} (ARG1).\\
    \bottomrule
    \end{tabular}
    \caption{\textsc{Open information extraction for example sentence.}
    Each proposition is an ordered list of arguments, separated by semicolons.
    Note the verbs (V) and agents (ARG0) in bold, highlighting the relation as ARG0 is performing V.}
    \Description{A table showing an example of open information extraction. The first row has an example sentence. The second row shows two relations extracted from the sentence, accompanied by their parts of speech tag.}
    \label{tab:action}
\end{table}

\subsubsection{Coreference Resolution}

To support \textbf{R1-R6} and \textbf{DG1}, we first need to detect all reference to the characters in the full text.
For example, consider the following two sentences: \textit{Gary is a journalist. He is a writer too.}
Here ``he'' refers to ``Gary.''
Identifying such co-references and clustering them correctly for a story is challenging as book length coreference resolution is an open problem in NLP~\cite{han2021fantasycoref}. 
However, this is critical for our case as extraction of traits such as actions and emotions will depend on correct identification of where the characters appear in the text.

To address this challenge, we tested three popular co-reference models: NeuralCoref~\cite{neuralcoref}, the AllenNLP coreference model~\cite{Lee2018HigherorderCR}, and the bookNLP co-reference model~\cite{DBLP:conf/lrec/BammanLM20}.
To evaluate the models, we split the books in our test dataset into chapters and then ran the models in each chapter to detect the co-reference clusters.
The chapter-wise split provides a manageable way to validate the coreference clusters.
We then calculated \textit{Error rate}, the average number of errors in the detected clusters for each model.
An error occurs when a mention has been incorrectly assigned to a cluster.
Overall, AllenNLP model had the lowest error rate, 1.5 (SD = 0.85).
NeuralCoref and bookNLP had on average 5 (SD = 2.1) and 4.1 (SD = 1) errors, respectively. 
AllenNLP also detected on average 2 more clusters than the other two models.
Thus, we decided to use AllenNLP for our system.

While AllenNLP is the most efficient among the three models, we noticed some mistakes by the AllenNLP model.
Such errors can lead to different traits being assigned to wrong characters.
AllenNLP also does not assign any name to a detected cluster.
Considering these factors, we adopt a human-centered approach to detect and validate the characters in the full text.
We used an interactive interface to validate, name, and merge the clusters (Figure~\ref{fig:annotate_interface}).  
 Finally, note that the coreference resolution already captures all mentions (i.e., presence) of each character, addressing \textbf{R1}.

\subsubsection{Action Detection}
To detect the actions, we need to identify verb and character pairs where characters are agents, performing the actions specified by the verbs.   
 Open Information Extraction (Open IE)~\cite{Stanovsky2018SupervisedOI} matches this requirement (\textbf{R2}).
Table~\ref{tab:action} shows an example sentence and two propositions extracted by Open IE.
We identify the agent (ARG0) and verb (V) pairs and search for the agent (ARG0) in coreference clusters for a possible match with a character. 
In the case of a positive match, we assign the verb to that character.
We manually annotated the actions for the lead characters in our test dataset and calculated the accuracy of this action detection method. It yielded a 95.7\% accuracy.

\subsubsection{Speech Detection}

To facilitate speech analysis, we extracted all direct quotes using a rule-based approach. 
For each direct quote, we look for self-referring words such as ``I'', ``me'', and ``mine'' and match with the coreference clusters.
In case of a match, we assign the character as the speaker of the speech.
For the speeches with no such match, we look for the closest verb just preceding or following the direct quote in the same sentence.
Here, we use part-of-speech tagging to determine if a given word is a verb.
Next, we find the corresponding subject for the closest verb using dependency parsing.
Here, the subject can be a noun or a pronoun.
In case of a pronoun, we use coreference resolution to determine the corresponding noun. 
This noun is attributed as the speaker of the direct quote. To evaluate this method, we again manually annotated the direct speeches or discourses from the lead characters in the test dataset.
This method yielded a 98\% accuracy on our test dataset.

 \begin{figure*}
    \centering
    \includegraphics[width=0.8\linewidth]{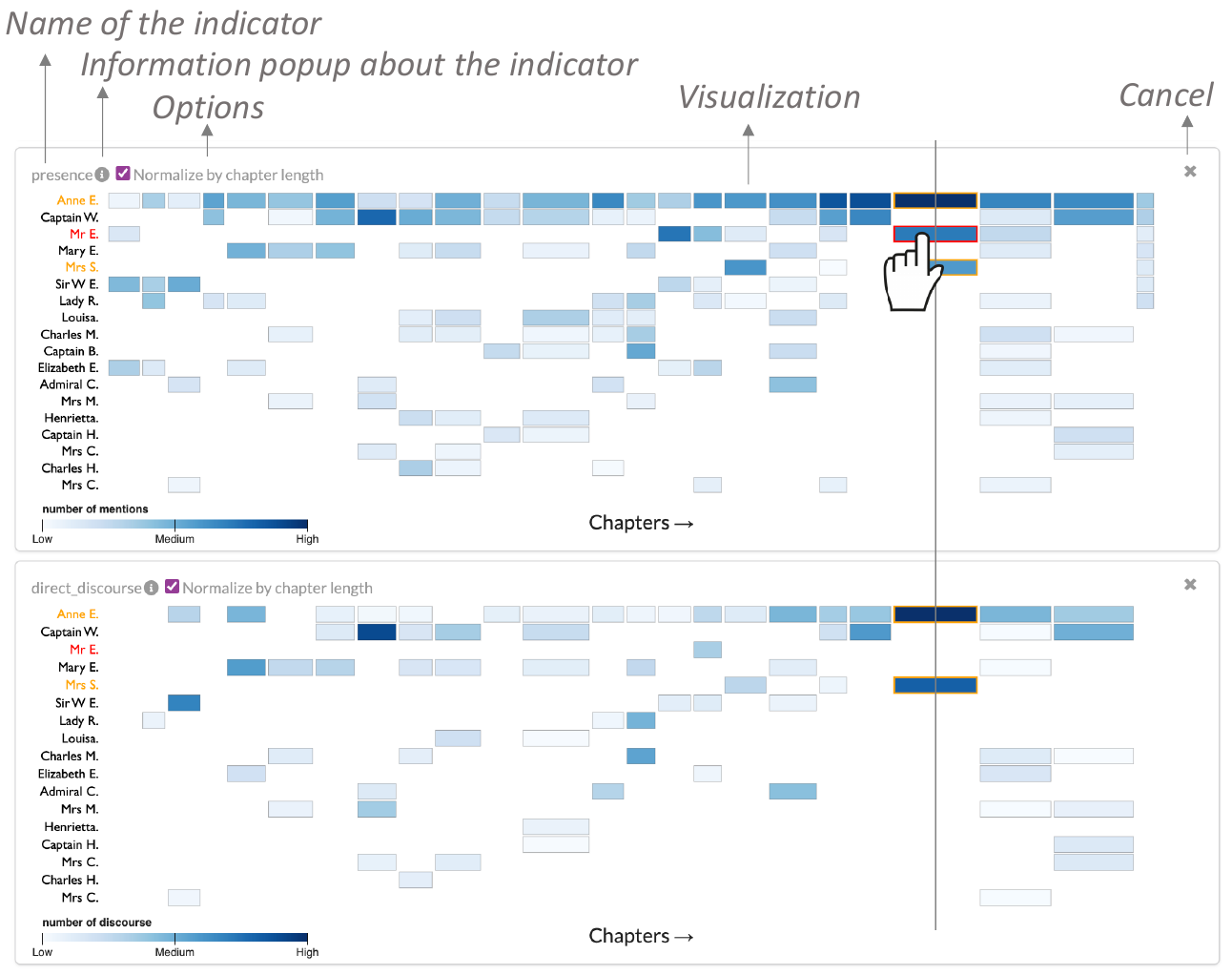}
    \caption{\textsc{Card view.}
    Each trait indicator is displayed in a card with homogeneous design.
    A card contains the \textit{name} of the indicator, \textit{information} about the indicator, additional \textit{controls} to revise the visualization, the \textit{visualization} representing the trait indicator, and an \textit{icon} to cancel/remove card.
    With this design, multiple cards can be aligned vertically.
    Here, we visualize two indicators in two cards: \textit{presence} and \textit{changes in actions}.
    The book displayed is \textit{Persuasion} (1818) by Jane Austen.
    }
    \label{fig:card}
    \Description{A figure showing a screenshot of Portrayal. Two rectangular boxes are aligned vertically. The first box shows the presence of the characters in the story with a collection of tiles. The leftmost tiles are from the first chapter, while the rightmost ones are from the last chapter. The tiles are color coded; the brighter the color, the more is the presence. The second box has similar visual style except it represents number of direct speech or discourses.}
\end{figure*}

\subsubsection{Direct Definition}

Adjectives work as direct definition to characters (\textbf{R4}).
To detect direct definitions, we first iterate through all words and identify adjectives using part-of-speech tagging.
Next, we try to identify the corresponding subject for each adjective.
To do that, we use dependency parsing and hop through different nodes in the tree from the adjective to a noun/pronoun.
This is achieved in two steps: (1) moving from the adjective to its parent node until we find a verb; and (2) moving from the verb to its subject.
In case any of these steps do succeed, i.e., we do not find a corresponding verb or subject, we ignore that adjective and move on.
In case the subject is a pronoun, we use coreference resolution to find its corresponding noun.
This entire process is repeated for all adjectives to find a mapping between an adjective and its subject.
Lastly, we aggregate all adjectives corresponding to each character.
This method was 91.2\% accurate on detecting adjectives for corresponding characters on our test dataset.

\subsubsection{Sentiment and Emotion Analysis}

We used sentiment analysis model from Allen NLP (90\% accuracy) to detect sentiment for each sentence (\textbf{R5}).
It predicts sentiment score on a binary continuous scale (-1 to +1).

We also used two pre-trained models for emotion detection.
The first model is Google's T5 base model~\cite{DBLP:journals/jmlr/RaffelSRLNMZLL20} fine tuned on a emotion detection dataset~\cite{DBLP:conf/emnlp/SaraviaLHWC18}.
It detects emotion on six categories: \textit{joy, sadness, love, anger, fear, and surprise}.
The second model is a RoBERTa base model~\cite{DBLP:journals/corr/abs-1907-11692} fine-tuned on the above emotion dataset.
Users have the choice of selecting any of the two models to use.

\subsubsection{Context and Environment Detection}

To extract contextual and environmental information, we used both coreference resolution and entity detection (\textbf{R6, R10}).
We noticed that, aside from characters, coreference resolution can also detect entities such as places and events with repeated mentions. 
Using the annotation interface (Figure~\ref{fig:annotate_interface}), we can label a cluster as a contextual tag (Figure~\ref{fig:annotate_interface}).
We further ran Spacy's entity recognition model to detect any missing places or time references.
Finally, writers often add small titles to the chapters that can work as contextual information.
We extract those using regular expression.

\subsubsection{Changes in Actions}

While action is an important trait, our formative study suggests writers and scholars also value changes in actions over time (\textbf{R8}).
To calculate change, we built upon Relative Norm Difference~\cite{DBLP:journals/pnas/GargSJZ18}, a metric that can measure difference between two sets of embedding vectors. 
Let $A_{ic}$ be the set of actions for character $c$ from chapter $i$ and $A_{(i-1)c}$ be the set of actions for character $c$ from chapter $i-1$.
Let $m_{ic}$ and $m_{(i-1)c}$ be the mean vectors for all the words in $A_{ic}$ and $A_{(i-1)c}$.
We define the change between $A_{ic}$ and $A_{(i-1)c}$ as $dist(A_{ic}, A_{(i-1)c}) = cosine(m_{ic}, m_{(i-1)c})$.
The range for this function is 0 to 1, where 0 means no change and 1 means the highest amount of change.

\subsubsection{Changes in Sentiment}

Change is also important for sentiment (\textbf{R8}).
\revise{However,  sentiment may fluctuate significantly on short text intervals (e.g., in a passage), making it difficult to see long-term patterns.}
Since sentiment is a continuous variable (between -1 and +1), we used moving average to smooth out short-term fluctuations and highlight longer-term trends and changes.
We used $n=5$ as window size for a full length story and $n=3$ for a short story.

\subsection{Visual Interface}

The visual interface of \toolname{} (Figure~\ref{fig:teaser}) is mainly divided into three parts: (a) Control Panel, (b) Visualization Panel, and (c) Text Editor.
We describe these parts below.
We include a companion \textbf{video} of \toolname{} in the supplement.
We also include a few design alternatives in the supplement that were explored in the earlier stage of this research.

The top panel of \toolname{} (Figure~\ref{fig:teaser}a) serves as a control panel.
It contains two dropdowns, one for selecting the trait indicators and the other for selecting the characters to visualize.
The ``Show Traits'' button updates the visualizations in the central panel based on the dropdown selections.
The optional ``Show Context'' checkbox lets users choose whether they want to see the context of a textual passage or not.
The visualization panel presents interactive visual representations of the trait indicators.
We describe the design of this panel below.

\subsubsection{Card Design}

To support \textbf{DG2}, we visualized each trait indicator as a separate \textit{card}, following the design of the popular UI component with the same name.\footnote{\url{https://www.nngroup.com/articles/cards-component/}}
All card views follow the same design.
In that way, a card works as a modular and homogeneous unit for our interface.
For example, Figure~\ref{fig:card} shows the basic components of a card: the name of the indicator, information about the indicator, control and filtering options for the indicator, a cancellation icon, and the main visual representation. 

\subsubsection{Timeline Representation}

Motivated by the dispersion plot commonly used in the humanities~\cite{jacobs2018gutenberg, DBLP:conf/ACMdis/HoqueGE22}, we used an occurrence matrix to represent trait indicators that depend on time (\textbf{R11, DG2}).
Figure~\ref{fig:card} shows an example of occurrence matrix-based timeline.
By default, the horizontal axis represents chapters in the story.
If there are no chapters, the axis defaults to one chapter only.
The vertical axis represents the characters.

We encode indicator values using a color scale.
For example, in Figure~\ref{fig:card} (top row), we encode the number of mentions for the characters in each chapter in the matrix cells with a linear color scale (i.e., the brighter the color of a cell, the more are the mentions for the character in the specific chapter).
Similarly, in Figure~\ref{fig:card} (bottom row), we encode the number of direct speeches for the characters.
Figure~\ref{fig:teaser} shows another example where three timelines---one for each trait indicators \textit{direct discourse or speech}, \textit{sentiment}, and \textit{emotion}---are aligned vertically.

\begin{figure*}[htb]
    \centering
    \includegraphics[width=0.9\linewidth]{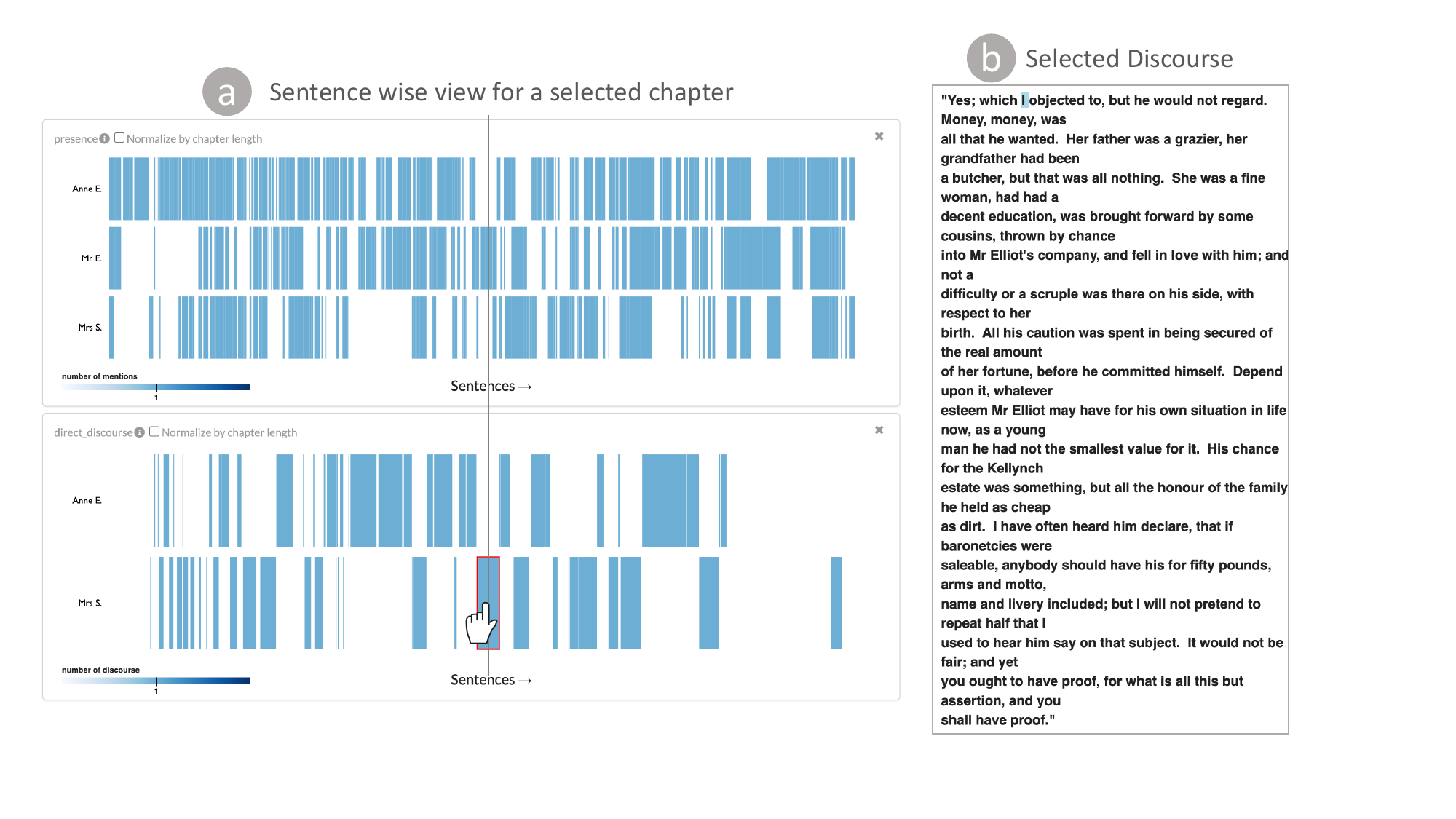}
    \caption{\textsc{Sentence view.}
    (a) When a user clicks on a chapter in the default view, we update the cards by showing all the sentences for that chapter.
    (b) A user can hover over a cell to see the particular sentence or discourse highlighted in the editor.}
    \Description{A figure showing a screenshot of Portrayal. Two rectangular boxes are placed vertically with an additional box placed on their side. The first box shows the presence of the characters in a selected chapter with a collection of tiles. The leftmost tiles are from the first sentence, while the rightmost ones are from the last sentence. The tiles are color coded; the brighter the color, the more is the presence. The second box has similar visual style except it represents number of direct speech or discourses. The third box shows a direct speech text in the text editor.}
    \label{fig:sentence_view}
\end{figure*}

\subsubsection{Interacting with the Timelines}

There are two ways to interact with the timelines: (a) mouse hover and (b) mouse click.
These interactions are designed to link the visualization with the text (\textbf{DG4}).
On hovering over a cell, we show the character and chapter name in a popup.
We also highlight the characters that co-occur with the selected character in the chapter.
This mechanism extends across multiple cards.
For example, in Figure~\ref{fig:card}, we hover over a cell for the character Mr.\ Elliot (shorthand Mr.\ E) in the presence card (from Jane Austen's \textit{Persuasion} (1818)).
We notice that there is no corresponding cell for Mr.\ Elliot in the direct discourse or speech view.
This suggests even though Mr.\ Elliot is highly mentioned in the chapter, he did not have any direct discourse.

To investigate further, a user can click on the cell.
In that case, the cards show the sentences of the chapter (Figure~\ref{fig:sentence_view}).
Interestingly, we see that there is a lot of direct discourses for Anne Elliot (shorthand Anne E.) and Mrs.\ Smith (Mrs. S.), but none for Mr.\ Elliot.
The explanation lies with the fact that Anne Elliot and Mrs.\ Smith are conversing about Mr.\ Elliot's past affairs and intentions.
Figure~\ref{fig:sentence_view}b shows a long speech from Mrs.\ Smith where she talks about Mr.\ Elliot's lust for money.
Note the colored background for the ``I'' in the first sentence which indicates the speaker and why our model thinks Mrs.\ Smith is the speaker in this case. 

\begin{figure*}[htb]
    \centering
    \includegraphics[width=0.9\linewidth]{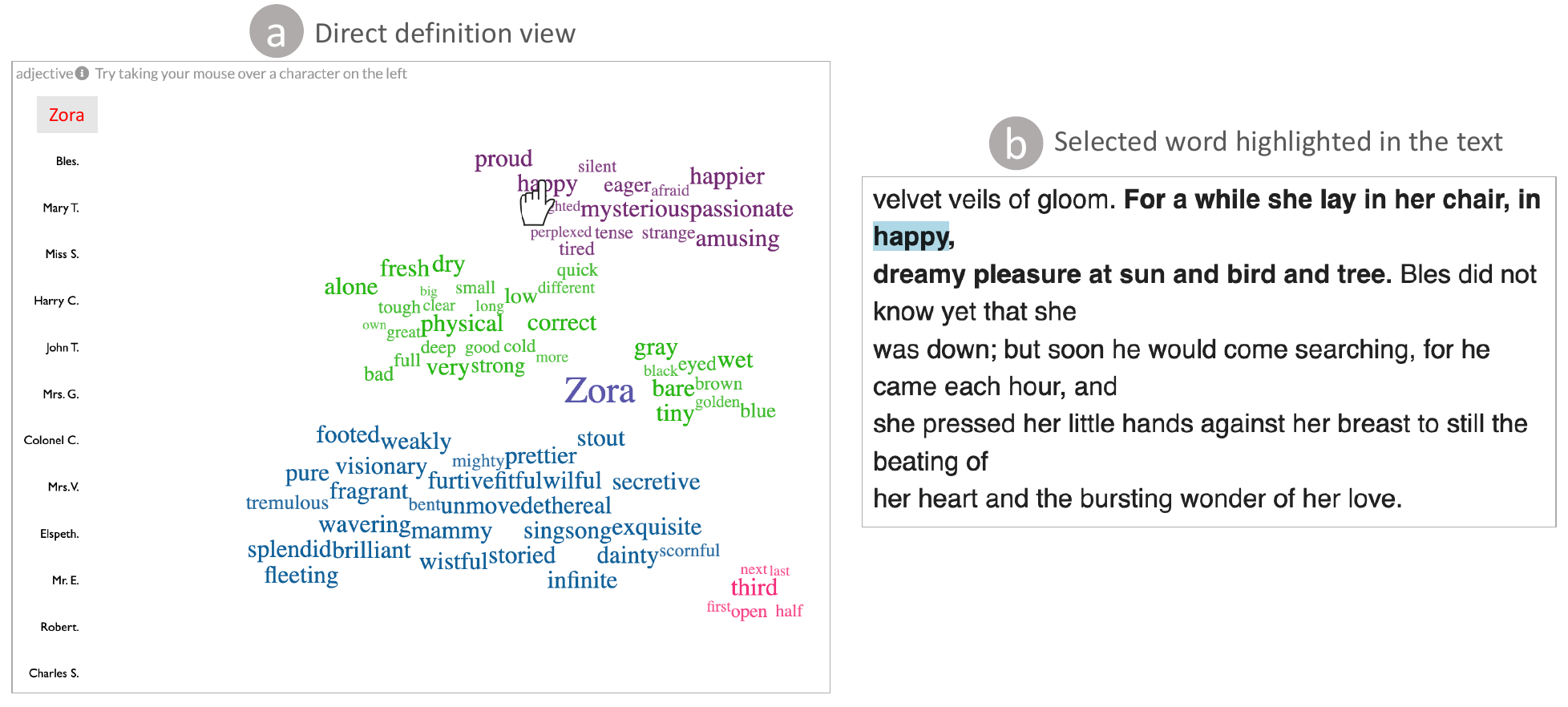}
    \caption{\textsc{Direct definition view.}
    (a) A user can hover over a character (on the left) to see a word zone representing sample adjectives and external appearances of that character.
    In this example, we see direct definition for Zora, one of the lead characters from the book \textit{The Quest of the Silver Fleece} (1911) by W.E.B. Du Bois.
    (b) The user can hover over a word to highlight the word in the text editor.}
    \Description{A figure with two sub-figures placed side by side. The left sub-figure shows a collection of words. The right figure shows a excerpt in the text editor, with a word highlighted in the text.}
    \label{fig:direct_definition}
\end{figure*}

\begin{figure*}[htb]
    \centering
    \includegraphics[width=0.9\linewidth]{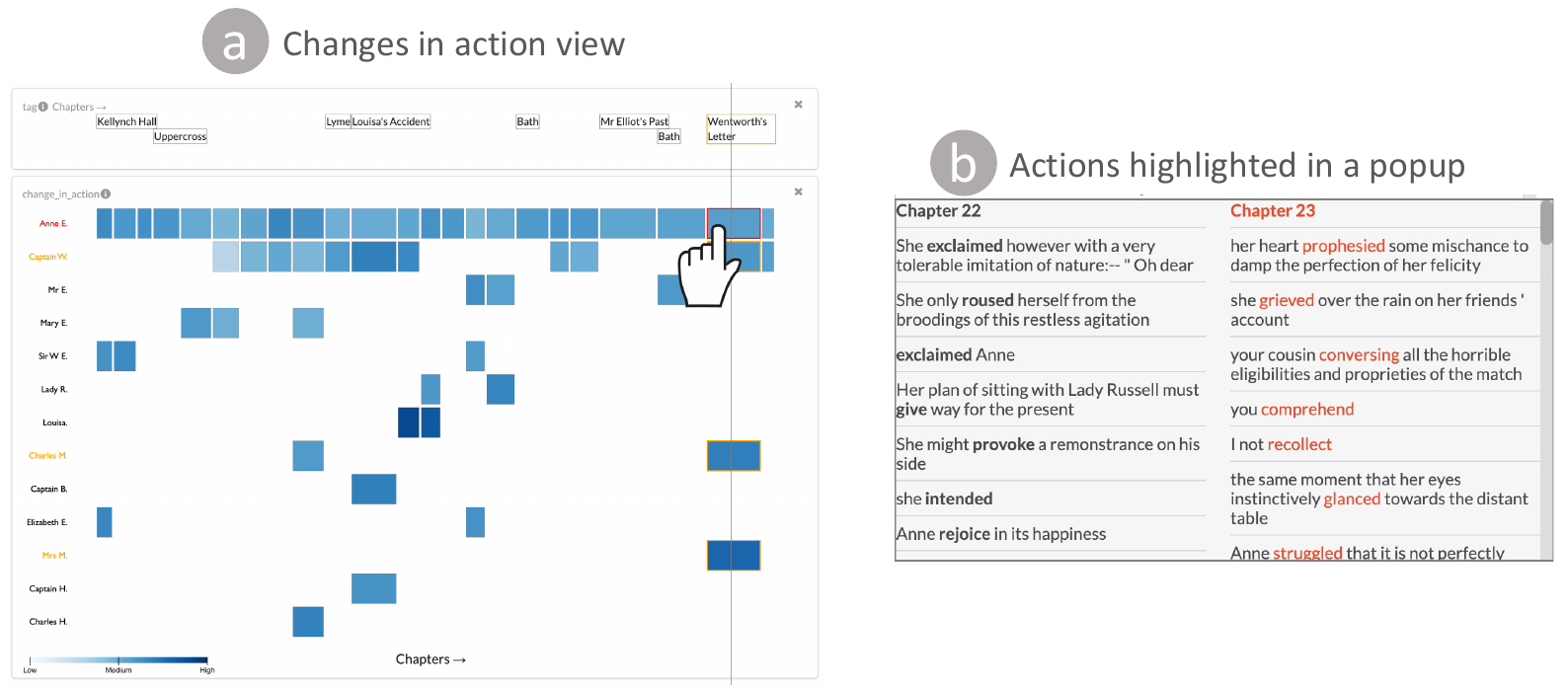}
    \caption{\textsc{Changes in action view.}
    (a) We encode the amount of changes for the actions from a chapter with respect to the previous chapter.
    (b) On hovering over a cell, we show the actions for the selected character (Anne Elliot) from the selected chapter (red) and previous chapter in a popup.
    The actions are sorted so that the most dissimilar pairs appear in the top.}
    \Description{A figure with two sub-figures placed side by side. The left sub-figure shows the changes of actions for the characters in a story. The leftmost tiles are from the first chapter, while the rightmost ones are from the last chapter. The tiles are color coded; the brighter the color, the more is the change. The right sub-figure shows two lists, one with actions from a chapter and the other with actions from a previous chapter.}
    \label{fig:changes_in_action}
\end{figure*}

\subsubsection{Actions and Direct Definition}

Four trait indicators (presence, direct discourse, sentiment, and emotion) in our interface use the occurrence matrix representation.
While we wanted to keep a homogeneous design across all visualizations, some trait indicators were not suitable for visualizing in a timeline.
More specifically, actions and direct definitions are mainly words and do not have a specific score attached to lay them out in a timeline. 
Therefore, we visualize them in word zones~\cite{hearst2019evaluation}.
Figure~\ref{fig:direct_definition} shows a word zone for the character Zora from the story \textit{The Quest of the Silver Fleece} (1911) by W.E.B.\ Du Bois.
To cluster the words into semantic groups, we perform $k$-means clustering on the embedding of the words.
Further, we identify the font weight for a word ($w$) for a character ($c$) as:  $weight(w, c) = tf(w,c) * (1/df(w))$, where $tf(w,c)$ is the frequency of $w$ for $c$ and $df(w)$ is the frequency of $w$ in the whole story.
It is essentially a normalized version of \textit{tf-idf}. 

Actions are also visualized in the same way.
However, based on \textbf{DG3}, we also wanted to visualize changes in the actions.
Since this a time-dependent indicator, we again use the matrix view to visualize it.
Figure~\ref{fig:changes_in_action} shows this representation for the story \textit{Persuasion}.
While we retain a similar representation, the interaction behaviour is slightly different for this indicator.
Because the number of actions can be large for a chapter, we rank and summarize the actions in a popup (Figure~\ref{fig:changes_in_action}b). 

\subsubsection{Contextual Information}

We show the contexts in a separate optional card (\textbf{DG5}).
The contexts align with the chapters in the timeline.
For example, Figure~\ref{fig:changes_in_action} shows the contexts on top of the changes in action card.
We also highlight the contexts when a user interacts with the timelines.
For space restriction and aligning the contexts with the cells, we can only show a limited number of contexts.  
To avoid overlaps among them, we use dynamic programming to find vertical positions of the texts whereas the horizontal positions are determined by the corresponding chapter.

\subsection{Implementation Notes}
We used Python, JavaScript, and D3~\cite{bostock2011d3} for managing back-end, front-end, and interactive visualization in Portrayal. We used SemanticUI and Bootstrap for styling various visual objects. 
We used Spacy~\cite{spacy} and AllenNLP~\cite{allennlp} for the NLP tasks.
The text editor in \toolname{} uses QuillJS~\cite{QuillJS} for a rich text editing functionality.
It is equipped with traditional formatting features such as selecting fonts, font sizes, font weights, etc., enabling the user to also write and modify content. The source code for \toolname{} is available in the supplement.

%% file: sections/7evaluation.tex
\section{User Study}
\label{sec:study}

We evaluated \toolname{} with a qualitative study involving literary scholars, creative writers, and participants who were both.
Our goal with this study was to understand how \toolname{} could help both audiences.

\begin{table*}[t!]
    \centering
    \begin{tabular}{lllp{2.23cm}p{6.8cm}c}
    \toprule
    \rowcolor{gray!10} 
    \textsc{\textbf{Id}} & \textsc{\textbf{Gender}} & \textsc{\textbf{Age}} & \textsc{\textbf{Profession}} & \textsc{\textbf{Experience}} & \textsc{\textbf{Yrs Exp}}\\ 
    \midrule
    P1 & Male & 25-34 & Literary scholar & BA and MA in English Literature, PhD candidate in communication & 10  \\
    
    \rowcolor{gray!10}
    P2 & Female & 25-34 & Literary scholar & Doctoral student focusing at European and African Literature (Anglophone and Francophone) & 10 \\
    
    P3 & Female & 18-24 & Literary scholar & BA and MA in English & 6 \\
    \rowcolor{gray!10}
    P4 & Male & 25-34 & Literary scholar & Doctoral student focusing on computation methods for understanding narrative structure & 6 \\
    
    \midrule
    P5 & Female & 25-34 & Creative writer & Novels (fiction/non-fiction), short stories, screenplays, poems, blogs, critiques, and fanfiction & 15 \\
    
    \rowcolor{gray!10}
    P6 & Male & 35-44 & Creative writer & Novels (fiction/non-fiction); short stories & 20 \\
    
    P7 & Male & 18-24 & Creative writer & Prose, poetry, short films, comedy sketches, creative essays; Creative writing minor; & 10 \\
    
    \rowcolor{gray!10}
    P8 & Female & 18-24 & Creative writer & Novels (fiction/non-fiction), short stories; Creative writing minor & 8 \\
    
    \midrule
    P9 & Male & 35-44 & Literary scholar \& creative writer & Doctoral student with interest in using text mining for contemporary Persian literature; published novelist & 18 \\
    
    \rowcolor{gray!10}
    P10 & Female & 25-34 & Literary scholar \& creative writer & MA in English and MFA in creative writing; short stories, novels & 15 \\
    
    P11 & Female & 45-54 & Literary scholar \& creative writer & BA and MA in English; creative writing minor; novels & 20  \\
    
    \rowcolor{gray!10}
    P12 & Male & 18-24 & Literary scholar \& creative writer & Doctoral student focusing on computation methods for understanding narrative structure; published short story writer & 5 \\
    
    \bottomrule
    \end{tabular}
    \caption{\textsc{Participant demographics.}
    Participants were recruited through our university's English department, literary center, and general mailing lists.
    All were required to have professional experience in literary scholarship and/or creative writing.
    }
    \Description{A table showing demographic information of twelve participants. The table has six columns: ID, Gender, Age, Profession, Experience, and Years of Experience.}
    \label{tab:participant}
\end{table*}

\subsection{Participants}

Similar to our formative study, we have three sets of participants---literary scholars (P1-P4), creative writers (P5-P8), and participants who were both (P9-P12).
We recruited our participants by advertising in the Department of English, creative writing program, and other mailing lists in our university.
Our inclusion criteria were similar to the formative study.
Three participants (P2, P5, P9) participated in our formative study; others were newly recruited.
Participant details are provided in Table~\ref{tab:participant}.

\subsection{Stories and Tasks}
\label{sec:tasks}

We anticipated that literary scholars would mostly be interested in analyzing well-known stories from literature.
While analyzing literature is a part of creative writing, we anticipated that writers would be most interested in analyzing their own stories. 
Thus, we wanted both groups of participants to analyze stories from literature using our tool, while writers should also analyze their own stories.
We preloaded four stories in our system for the study.
We consulted a literary scholar, a professor of English at our university, and a co-author of this paper to select the stories.
Our inclusion criteria involved trying to balance three sometimes conflicting considerations: works that are (1) likely to be well-known to literary scholars and creative writers, (2) are representative of authors from diverse backgrounds, and (3) are publicly available in Project Gutenberg.
Based on these criteria, we chose the following stories: 

\begin{enumerate}

    \item \textit{Persuasion} (1818) by Jane Austen, adapted into several movies, most recently in the 2022 eponymous Netflix one.
    \item \textit{The Yellow Wallpaper} (1892) by Charlotte Perkins Gilman, regarded as an important early work of American feminist literature.
    \item \textit{The Quest of the Silver Fleece} (1911) by W.\ E.\ B.\ Du Bois, a famous African-American sociologist, activist, and historian most known for his work against slavery and racism.
    Interestingly, Du Bois also pioneered visualization work in the early 1900~\cite{DBLP:conf/chi/KarduniWCD20}.
    \item \textit{Alice in Wonderland} (1865) by Lewis Carroll, a well-known children's story.
\end{enumerate}

For each of the stories, we designed a few multiple-choice questions to aid in the evaluation and help participants get acquainted with the tool.
We provide the full list of questions as supplementary material.
Here are a few example questions for the story \textit{Persuasion} below:

\begin{enumerate}
    \item According to the ``Presence'' feature, which characters were present in the first chapter?
    \item According to the ``Sentiment'' and ``Emotion'' feature, which character faced the most emotional changes in Chapter 23?
    \item According to the ``changes in actions'' feature, which chapter saw the highest changes in actions for Captain Wentworth?
\end{enumerate}


The questions were designed to guide participants to explore different features of the tool, not to test their comprehension of the stories.
Finally, we asked participants who were writers (\textbf{P5-P12}) to provide us a draft of one of their unpublished stories.
We then uploaded each story into the tool.
Each writer only had access to their own story, not stories written by other participants. All study materials are available in our \textcolor{cyan}{\href{https://osf.io/2pqax/}{OSF repository}} and supplement.

\begin{figure*}
    \centering
    \includegraphics[width=0.9\linewidth]{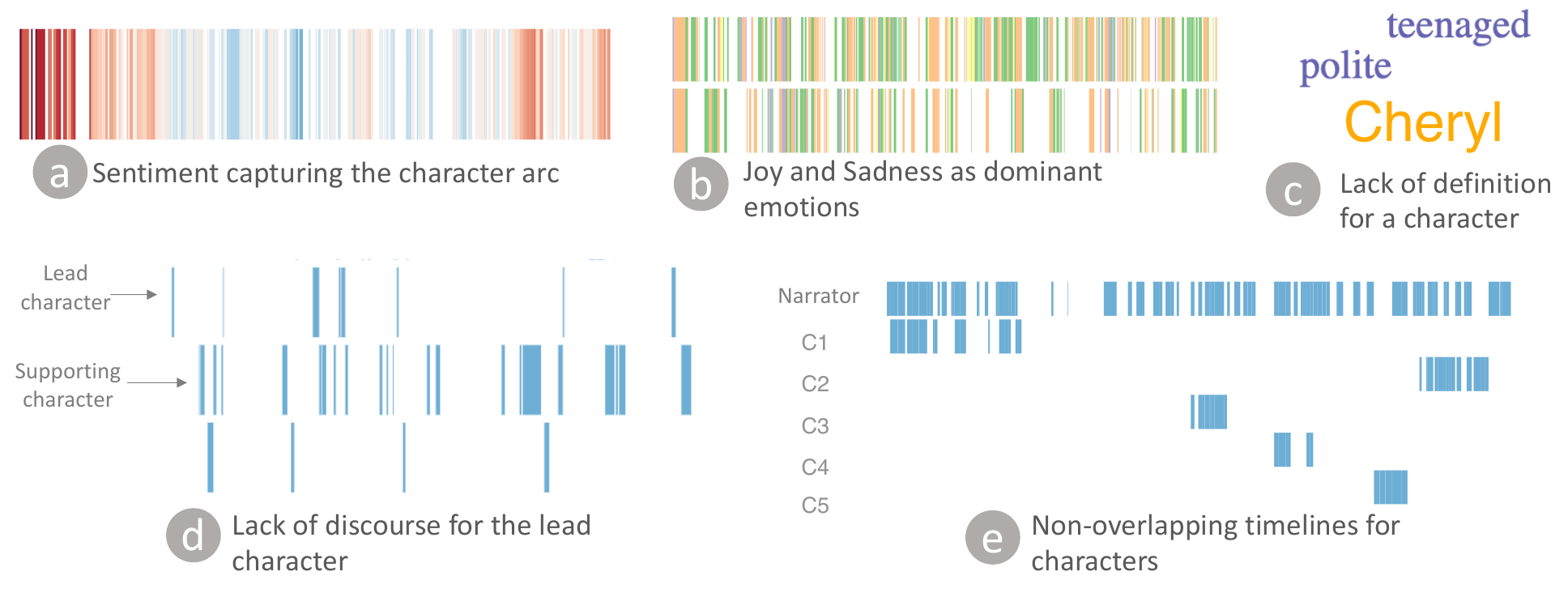}
    \caption{\textsc{Examples of actionable insights found by writers using \toolname{}}.
    (a) P11 used the smooth option in sentiment to see gradual change in sentiment for the lead character in their story.
    They were reassured to see it matched their intended arc for the character (negative $\rightarrow$ neutral $\rightarrow$ negative).
    (b) P12 was surprised to see joy and sadness interleaving for two characters in their story.
    They expected sadness to be the prominent emotion.
    (c) P10 found the character named Cheryl (pseudonym) to be underdeveloped by looking at just two direct definition for that character.
    (d) P6 discovered that the lead character did not have enough direct discourse as they would have liked compared to the supporting character.
    (e) P7 found that the characters did not interact with each other to their liking in the story.}
    \Description{Five images showing insights gathered by writers in the study. The first row has three images placed side by side. The first two images have colors tiles. The third image has a collection of words. The second row has two images placed side by side. Both images have a collection tiles in a timeline, placed left to right.}
    \label{fig:writer_insights}
\end{figure*}
\subsection{Procedure}

The study was conducted over Zoom and was divided into three parts:
(1) initial meeting with participants;
(2) deployment of the tool with participants;
and (3) post-study interview and feedback meeting.
Details on the parts follow.

\textbf{Initial meeting with participants (30 minutes).}
This was an introductory meeting between the participants and the study administrator (the first author of this paper).
After informed consent, the administrator introduced the tool to the participants with a live tutorial on a training story.
Participants then used the tool with guidance from the administrator.
We encouraged participants to ask questions at this stage.
Once participants felt comfortable with the tool, we provided them with a document that contained the link to the tool, links to the questions for each story, and the tool's documentation.
This document is provided in the supplement. 

\textbf{Deployment of the tool with participants (1 week).}
At this stage, participants used the tool independently for a week without any interference from the research team.
We instructed participants to use the tool for at least 1 hour on their own.
Literary scholars answered the questions for at least two preloaded stories, while creative writers answered the questions for one preloaded story and then analyzed their own story.
Finally, we encouraged participants to follow their intuition and own work process.
Participants were requested to reach out to the research team if they had any problems with \toolname{}; however, none did.

\textbf{Post-study interview and feedback meeting (30 minutes).}
Finally, we met the participants after one week and conducted semi-structured interviews to understand their experience with \toolname{}.
We include the semi-structured questionnaire in the supplement.
Participants also rated the tool on six subjective metrics on a Likert scale ranging from 1 (strongly disagree) to 7 (strongly agree).
At the end of this session, each participant received a \$40 gift card.

\subsection{Analysis}

Similar to the formative study, we created an anonymized transcript from the recorded audio for each interview.
The first author of this paper open-coded the transcripts.
A code was generated by summarizing relevant phrases or sentences from the transcripts with a short descriptive text.
The first author then conducted a thematic analysis~\cite{braun2006using} to group related codes into themes.
A second author independently validated the themes, looking for raw evidence in the interview transcripts.
Finally, the codes and themes were revised through discussion with the full author team.

\subsection{Study Design Rationale}

\subsubsection{Rationale for Qualitative Study}

We decided a qualitative study is best suited to evaluate \toolname{} for three reasons. 
First, creative works such as fiction writing and analysis typically do not adhere to predefined structures and largely depend on artists' styles and idiosyncrasies.
Thus, there is no objective measurement of efficiency for such tasks, which quantitative studies often try to measure.
Secondly, we anticipated that writers and scholars will not be comfortable if we put them into analytic tasks in a controlled environment for a limited time.
Thus, we wanted participants to use the tool in a familiar environment for an extended period, with the freedom to adjust to the tool at their will.
Finally, there are currently no analytic tools that focus on characterization against which \toolname{} can be quantitatively compared.
Although our work is built upon several prior works (Section~\ref{sec:design}), none of them were designed to understand characterization.
A logical baseline is a simple text editor without any analytic support.
However, given that our participants are professional writers and scholars, we believe they are already in a position to compare \toolname{} with a simple text editor. 

\subsubsection{Rationale for Study Tasks}

Aside from the guiding multiple choice questions (Section~\ref{sec:tasks}), we did not provide any explicit task list to the participants but instead encouraged them to follow their intuition and own creative work process.
This decision is again motivated by allowing participants to practice their artistic freedom at will.
Another critical decision is whether or not we should ask writers to write a story using our tool.
Writing a narrative fiction takes significant planning and effort, and is likely to be a very long and involved process.
We would require long time commitments from several writers.
We felt that this was impractical at this stage of the research project.
Thus, we instead decided to ask writers to analyze one of their work-in-progress drafts. 
This allowed us to understand what feedback \toolname{} can provide to writers during revising, one of the critical stages of writing~\cite{du-etal-2022-understanding-iterative}.
Additionally, writers provided feedback on how they would use the tool during writing from their personal experience.

\subsection{Results}

Here we present the findings from the interviews.
The findings relate to how \toolname{} augmented creative writing and literary analysis as well as the usability and limitations of the tool.

\subsubsection{Augmenting Creative Writing with \toolname{}}

Writers in our study did not report using any analytical support before \toolname{}{}.
They primarily relied on critical reading, multiple rounds of editing, and obtaining feedback from peers, friends, and publishers to revise their writing.
With the first use of any analytical support, participants shared several insights they gathered from \toolname{}, some reassuring, while others were surprising. 
Figure~\ref{fig:writer_insights} presents some of these insights.
We further present the themes that emerged from the post-study interviews below.

\textbf{Creating Dynamic Characters and Scenes.}
A critical challenge for writers is to create dynamic characters and scenes that engage readers.
Participants thought the indicators supported by our tool helped them in this regard (P5-12), \revise{validating our first design goal (DG1)}.
\revise{ We also noted that the use of multiple views helped writers analyze multiple indicators together (DG2).}
For instance, P10 mentioned that \textit{\textbf{sentiment}} and \textit{\textbf{emotion}} are helpful for creating characters that go through many emotional changes.

\begin{quote}
    \texttt{``I think sentiment and emotion are particularly useful for characterization.
    A good rule for characterization that I follow is having multifaceted characters that can basically show a variety of emotions and feelings.
    I want to have dynamic characters which are at some point optimistic but at other point is pessimistic, unless I want to have static characters.
    Using the tool, I can see parts where I want to make changes to achieve that goal.''} (P10)
\end{quote}

Sentiment and emotions are also helpful for creating scenes where characters with opposite emotions interact.
In P11's opinion, such scenes create ``tension'' and make the story more ``engaging''.

\begin{quote}
    \texttt{``One thing I try to follow, not that I achieve that all the time, is to have characters with different emotions interact.
    For example, interactions between a joyful character and a sad one may help create tensions in the story.
    The tool helped me validate that in my story.''} (P11)
\end{quote}

Another set of participants found \textit{\textbf{presence}} and \textit{\textbf{direct speech}} to be most helpful for creating dynamic characters and scenes.
For example, P8 mentioned that the tool helped them validate Bakhtin's Polyphonic Theory.

\begin{quote}
    \texttt{``If you want to create a polyphonic novel, several characters should have voices instead of just one.
    The alignment of timelines (in presence) is really helpful to see whether you have enough characters in the scenes and chapters.
    Discourse (speech) analysis enhances that by showing how you manage the voices of the characters.''} (P8)
\end{quote}

Finally, \textbf{\textit{actions}} and \textbf{\textit{direct definitions}} are helpful for determining whether a character has developed or not.
For example, Figure~\ref{fig:writer_insights}c shows one such example where the lack of direct definition helped P10 identify an underdeveloped character. 


\textbf{Capturing Character Arcs.}
Participants found that \toolname{} was able to capture character arcs (P5, P9, P12).
A character arc is a transformation or inner journey of a character over the course of the story.
Changes in action and sentiment can partially capture it, \revise{thus validating DG3}.
For example, P5 and P9 said: 

\begin{quote}
    \texttt{``I think the change in action is one of the most useful for sure.
    Just being able to see the way actions jump from chapter to chapter is super interesting.
    It would be really useful for thinking about a character arc.''} (P5)
\end{quote}

\begin{quote}
    \texttt{``The smooth option in the sentiment view was awesome.
    It perfectly captured the arc for my lead character!''} (P9)
\end{quote}

\textbf{Identifying unconscious bias.}
We noticed \toolname{} helped writers expose their unconscious bias towards a group of characters with a similar social identity (P6-7, P9, P11).
Writing a story, even a short one, takes significant effort and often requires multiple rounds of revising.
While reviewing a draft, \toolname{} can be helpful to reassure intended and unintended characterization for a group of characters. 
Participants found the presence and discourse indicators to be most beneficial for that.
We note that this finding is in line with the findings of DramatVis Personae~\cite{DBLP:conf/ACMdis/HoqueGE22}.
For example, P9 and P6 said: 

\begin{quote}
    \texttt{``A creative writer gets an objective perspective on which characters they give more presence and dialogue to in a story.
    This could help writers address their own subconscious biases as they could see it laid out for them---what type of characters they give priority to and what characters they let interact.''} (P9)
\end{quote}

\begin{quote}
    \texttt{``I want to see whether the female characters in my novel have had enough space for expression.
    But when I am writing, I am not consciously thinking about all of this.
    I am just writing and writing and writing.
    And then, when I review what I have written, I can see whether my female characters have the right exposure.''} (P6)
\end{quote}

\subsubsection{Augmenting Literary Scholarship with \toolname{}}

All scholars in our study reported close reading as the primary way to analyze a literary work.
Four scholars (P2, P4, P9, P12) reported having hands-on experience with text mining and NLP for distant reading.
However, none reported using any special purpose tool for close reading or analyzing characters.
We report how \toolname{} augmented literary analysis below, based on the themes that emerged from the interviews.

\textbf{Balancing Subjectivity and Objectivity.}
The indicators in \toolname{} will help scholars to support their literary argument with concrete examples and avoid unconscious predilection toward an argument (P1-3, P9) \revise{(DG1)}.
Scholars typically support their argument with a limited number of examples, which is susceptible to missing out on examples that oppose their argument.
For example, P2 and P9 said: 

\begin{quote}
    \texttt{``This tool will help scholars such as myself balance our subjectivity with the objective information provided in the tool.
    If I did not have this tool, I am only analyzing maybe three actions of a character.
    But, there may be one hundred and eighty-seven actions that do not support my argument, and I can unconsciously turn a blind eye to them.
    The tool removes this kind of blind subjectivity for a scholar and makes you look at the text of what the author put forward.''} (P2)
\end{quote}

\begin{quote}
    \texttt{``It has been very challenging since the beginning for literary analysis that we say this character is too pessimistic.
    But how do we support it?
    We hugely rely on sampling.
    We say in these three examples, he shows negative emotions.
    But here, you have more concrete evidence for the claims that you make.''} (P9)
\end{quote}

On the flip side, participants suggested that it is possible that scholars can get over-dependent on the tool and lose their subjective intuition (P1-2).
However, participants suggested this is unlikely since scholars are trained to exercise their literary knowledge and will likely find a balance between subjectivity and objectivity.

\textbf{Bridging Close Reading with Computational Support.}
\toolname{} provides a bird's eye view over a story (P3) and helps scholars drill down in a top-down approach for close reading (P1-P4, P10-12).
Close reading is an integral part of literary analysis.
However, participants mentioned that it requires enormous effort and time to annotate text.
\revise{The links between text and visualization reduce workload for scholars (DG4).}
P10 and P12 said:

\begin{quote}
    \texttt{``Most digital humanities tools (e.g., Voyant Tools) are designed for big data (distant reading)
    But this one helps with the close readings because it breaks down every character, every chapter, and even every sentence.
    Moreover, it starts with an overview of the chapters and helps me drill down the chapters of interest.
    This was a completely new and amazing experience for me.''} (P10)
\end{quote}

\begin{quote}
    \texttt{``The best feature is the micro level link between visualization and text.
    I think I will be blind without it!''} (P12)
\end{quote}

\textbf{Facilitating Literary Conversation and Debates.}
Participants thought subjective indicators such as sentiment and emotion that depend on the interpretation of the AI model are more suitable for literary conversation and debates (P2-4, P9).
During the study, several participants (P2-4, P9) organized the indicators into two categories:
(1) Objective indicators such as presence, speech, and actions that are directly present in the text; and
(2) Subjective indicators such as sentiment and emotion that depend on the AI model's interpretation. 
 
Participants suggested that the amount of agreement and disagreement with the subjective indicators will vary from scholar to scholar.
This is in line with the current practices in literature where scholars can interpret the same text in different ways.
This process makes these indicators a good candidate for literary conversation and debates.
 
\begin{quote}
    \texttt{``Even in literary classes, when we used to do analysis, we had debates over what is the sentiment here, what is the emotion this character is experiencing.
    The tool could be a jumping-off point where you could say like, why would the tool pick a negative sentiment here? 
    Let's talk about keywords here.''} (P3)
\end{quote}
 
\begin{quote}
    \texttt{``It is interesting to see how the AI is inferring sentiment.
    I think this could be a talking point for people who want to use technology to analyze stories.
    You can find some controversial examples from the tool.''} (P4)
\end{quote}

\revise{P4 further mentioned that such debates and conversation are generally constructive.
However, it is possible that scholars who use \toolname{} might receive criticism, as the use of AI is a polarized topic in the scholarly community.}

\textbf{Writing Literary Essays with \toolname{}.}
All scholars in our study had already graduated from college.
However, several participants (P1-4, P10, P12) reminisced about their college days during their studies when they had been writing literary and term final essays. 
These participants felt that the tool would be useful for that. For example, P1 said:

\begin{quote}
    \texttt{``When I am trying to write an essay, what's usually the hardest thing to do is find like oh when did the character say this or did that, or when did this character show up or find passages from the book and connect them.
    Being able to do this with the tool will honestly be life-changing for students.''} (P1)
\end{quote}





\begin{figure*}
    \centering
    \includegraphics[width=0.8\linewidth]{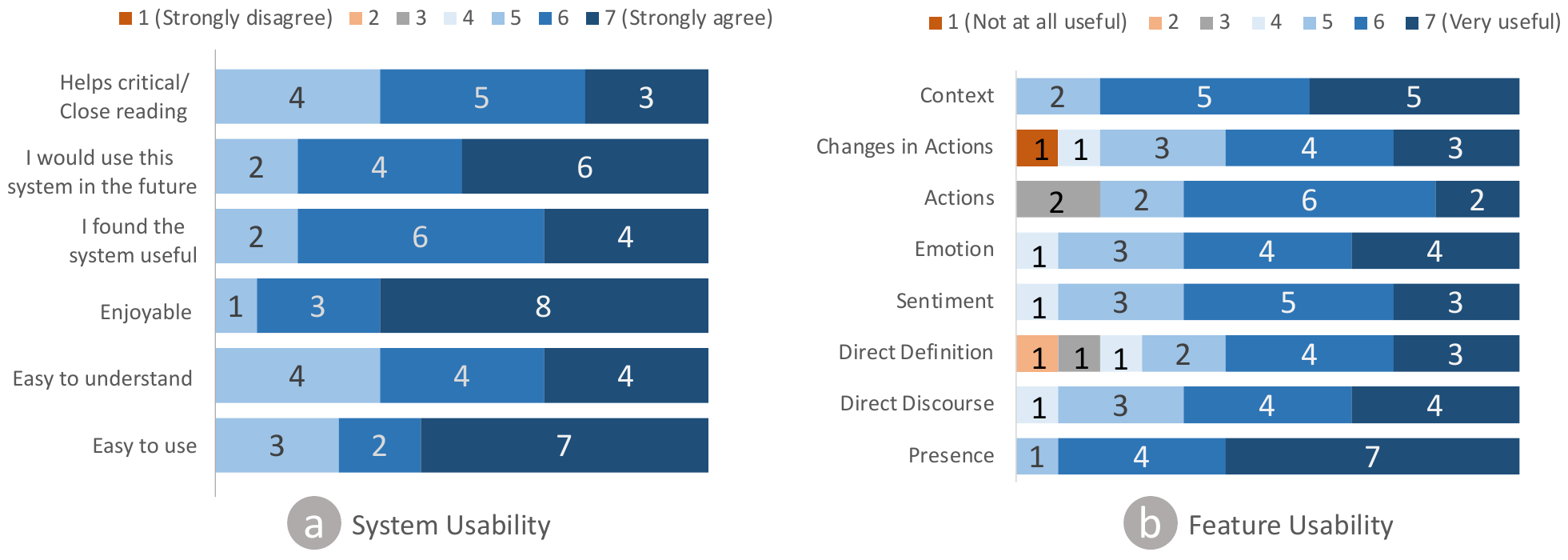}
    \caption{\textsc{Usability ratings.}
    (a) System usability ratings provided by participants on a scale from 1 (strongly disagree) to 7 (strongly agree).
    (b) Trait indicator usability ratings on a scale from 1 (no at all useful) to 7 (very useful).
    The numbers in the bars indicate the number of participants voted for the particular rating.}
    \Description{Two images showing usability ratings provided by participants. The images use colored tiles to show the level of ratings, from 1 to 7. There are numbers encoded in the cells, indicating the number of people voted for each category.}
    \label{fig:usability}
\end{figure*}

\begin{table*}[t]
    \centering
    \begin{tabular}{p{5.5cm}cccccc}
    \toprule
    \rowcolor{gray!10} 
    \textsc{\textbf{Support}} & \multicolumn{6}{c}{\textsc{\textbf{Trait Indicator}}} \\
    
    & \textsc{\textbf{Presence}} & \textsc{\textbf{Discourse}} & \textsc{\textbf{Sentiment}} & \textsc{\textbf{Emotion}} & \textsc{\textbf{Action}} & \textsc{\textbf{Direct Definition}} \\
    \midrule
    
    Evaluate engagement and dynamic nature of characters* & \checkmark & \checkmark & \checkmark & \checkmark & \checkmark & \checkmark \\
    \rowcolor{gray!10} 
    Evaluating scene and chapter structure & \checkmark & \checkmark & \checkmark & & &  \\
    Evaluate character arcs* & & & \checkmark & \checkmark & \checkmark &  \\
    \rowcolor{gray!10}
    Evaluate character prioritization and bias & \checkmark & \checkmark & & & \checkmark & \checkmark \\
    \midrule

    Indexing and annotating text & \checkmark & \checkmark & & & \checkmark & \checkmark \\
    \rowcolor{gray!10}
    Finding and searching for patterns and examples (close reading) & \checkmark & \checkmark & \checkmark  & & \checkmark & \checkmark \\
    Literary conversation and debates & \checkmark & \checkmark & \checkmark  & \checkmark & \checkmark & \checkmark \\
    Writing literary essays & \checkmark & \checkmark & \checkmark  & \checkmark & \checkmark & \checkmark \\
    
    \bottomrule
    \end{tabular}
    \caption{\revise{\textsc{Use cases for the trait indicators.} The first four rows show use cases identified for creative writing. The last four rows show use cases identified for literary analysis. The \checkmark encodes which indicators are useful for a specific use case. The user study informed the use cases. *Changes in sentiment and actions are useful for these use cases although they are not listed as separate traits in the table.}}
    \Description{A table showing different use cases identified from the user study. The first column shows the use cases. The next six columns are the trait indicators: presence, discourse, sentiment, emotion, action, direct definition. A cell contains a check-mark if the trait indicator is useful for the specific use case. }
    \label{tab:summary}
\end{table*}

\subsubsection{Usability}

All participants admired the visual design of the interface.
Participants repeatedly used praises such as ``cool'', ``neat'', ``not overwhelming'', ``fluid'', ``easy to use'', and ``easy to get going'' to describe the system's usability.
This overall positive experience was reflected in the post-study usability ratings provided by the participants (Figure~\ref{fig:usability}). 
Across all system usability categories, there were no ratings less than 5 (Figure~\ref{fig:usability}a).
Similarly, participants rated most trait indicators highly in terms of how useful they are (Figure~\ref{fig:usability}b). 

\revise{One usability issue was the lack of a timeline for the word clouds (direct definition and actions).
For full length stories, P6 found word clouds to be difficult to use for finding patterns.
P6 suggested organizing the words in a timeline for better interpretability. 
Another usability issue was the long list of actions in the ``Changes in Actions'' view (Figure~\ref{fig:changes_in_action}b).
P1 found it difficult and time-consuming to read through the list.
P1's negative experience was reflected on the rating of 1 for this view (Figure~\ref{fig:usability}b-second row).
P9 found the color scheme in the Emotion view difficult to understand and wished for a smooth option similar to the one we have for Sentiment.}

\revise{Finally, participants reported that they frequently used pairs of indicators using our multiple view design.
However, use of three indicators was rare, slightly diminishing the effectiveness of DG2.
The increased attention required to analyze three indicators together  have discouraged users in such scenarios.}

\subsubsection{\revise{Errors, Limitations, and Suggestions for the NLP Pipeline}}

While the experience with the tool was mostly positive for the participants, there were a few concerns and limitations outlined by the participants.
We discuss them below.

\textbf{Lack of explanation for the AI models.}
Several participants inquired about the underlying models that were used in \toolname{} (P2, P6-8, P12).
During the study, we provided participants with a description of the models and links to the original models.
However, most of our models were based on deep learning, making them black boxes with very little explanation of how they work internally. 
This clearly created a lack of trust among some participants, especially for the subjective indicators (sentiment and emotion).
We discuss ways to mitigate this concern in Section~\ref{sec:discussion}.

\textbf{Errors in NLP Pipeline.}
Participants found a few errors with the \revise{co-reference} resolution model.
It was expected since these models are not perfect.
However, participants suggested that it did not impact their experience negatively as they were surprised with the accuracy of the \revise{co-reference} model, especially in complex sentences.
Or, as P1 put it: \texttt{``The tool could determine mentions that I might have missed when close reading.''} (P1)

    

Additionally, participants found a few adjectives that were wrongly determined as the direct definition of the characters.
Due to some ambiguous cases, our rule-based model could not discern and can be resolved easily with additional rules.

\subsubsection{Improvements.}
Participants requested a few new features in the tool.
First, P9 noted that comparing trait indicators in a corpus would be interesting for facilitating distant reading.
Similarly, P2 and P12 requested features that can compare two versions of the same story to compare their revised version with the original one.
P11 suggested including ``point of view'' as a feature.
For example, writers often narrate the story from the point of view of a narrator or multiple characters.
Finally, P6 requested the ability to show conversations more clearly.
We include all speech of the characters in the direct discourse view.
However, we do not identify or show who is spoken to in the speech, only who is speaking.
This information can show how writers manage conversations more clearly.

%% file: sections/8discussion.tex
\section{Discussion}
\label{sec:discussion}

In this section, we discuss the broader impact, limitations, and future research directions of \toolname{} below.

\subsection{\revise{Reflecting on the User Study}}

\revise{We believe \toolname{} has shown to successfully augment creative writing and literary analysis.
Table~\ref{tab:summary} summarizes the use cases identified from the user study. 
It shows the diverse creative writing and analytic tasks that can be performed using \toolname{}. Participants used the trait indicators for diverse use cases in various combinations.}

\revise{Another interesting finding is the contrast among the use cases for writers and scholars.
Writers primarily focused on developing characters and scenes.
Surprisingly, scholars went beyond character analysis and were able to integrate \toolname{} in their broader workflow (e.g., close reading, literary essays, character studies).
This indicates the generalizability of \toolname{} and validates our decision to support both writers and scholars in the same tool.}

\subsection{\revise{Design Implications for Future Creativity Support Tools}}

\textbf{NLP as an Analytic Partner.}
Our success with \toolname{} suggests it has the potential to rekindle research on using NLP as an \textit{analytic} assistant.
This is in contrast with recent work on creative writing,  which tend to focus on \textit{generative} \revise{applications}~\cite{talebrush, lee2022coauthor, yuan2022wordcraft}.
This trend is not surprising since generative \revise{applications} are an exciting new frontier in which the machine acts as a co-writer and can improve creative writing.
However, we believe NLP as an assistant---when applied to reveal hidden patterns from stories rather than creating stories---still has powerful applications to offer. 
This work showed one example of such applications.
We believe that bridging the two directions---NLP both for story generation and for analytic assistance---is the right way to cement NLP's importance in creative writing and literary analysis.

\textbf{Implications for Visualization in Creativity Support.}
While providing computational lenses on text is common in creativity support tools, they are often limited to one or two lenses~\cite{DBLP:conf/ACMdis/HoqueGE22, DBLP:conf/uist/DangBLB22, sterman2020interacting}.
We believe the use of interactive visualization has enabled us to couple several computational lenses (i.e., models) together in \toolname{}.
Different models in \toolname{} work as different computational lenses to look at the story.
Each model provides a unique perspective to writers and scholars, and their combination is yet another perspective.
We believe our work can motivate researchers to use interactive visualization as a medium to represent complex narrative components in the future.



\subsection{\revise{Opportunities for Interactive Machine Learning in Creativity Support Tools}}

\textbf{Integrating XAI into Creativity Support.}
A concern among our participants was the lack of explanation of the AI models---a well-known problem in machine learning.
The entire field of \textit{explainable AI (XAI)} is dedicated to addressing this problem. We envision integrating interpretable models~\cite{DBLP:conf/nips/KimRS14, angwin2016machine} to mitigate this concern.
For example, aside from our current sentiment analysis model, we can integrate VADER~\cite{DBLP:conf/icwsm/HuttoG14}, an interpretable model for sentiment analysis in our tool.
It uses a dictionary of words, where each word has a sentiment score between -4 to +4.
The sentiment score of a sentence is calculated by summing up the sentiment score of each word and then normalizing it to the -1 to +1 range.
By investigating the sentiment scores for individual words, a user can understand the reason behind the sentiment score of the sentence.
We could integrate such models into our tool without making any major changes. 

A critical consideration here is that interpretable models tend to be less accurate than their black box (deep learning) counterparts (i.e., the accuracy interpretability trade-off~\cite{DBLP:journals/tvcg/HoqueM22}).
We believe the right way to move forward is to integrate both interpretable and black box models and allow users to choose a model of their choice.
Additionally, black box models can use a post-hoc explanation module such as LIME~\cite{DBLP:conf/kdd/Ribeiro0G16} or SHAP~\cite{DBLP:conf/nips/LundbergL17} to improve interpretability.
Both of these methods can measure feature importance (word importance) for any black box model.
We can visualize the feature importance in bar charts~\cite{lundberg2018explainable}, allowing users to understand the black box models.
Given the interactive visual nature of our tool, it will not be difficult to integrate such bar charts into our tool.

\textbf{Human-in-the-loop Co-reference Resolution.}
Book-length co-reference resolution was a major challenge for our work as it is still an unsolved problem in NLP~\cite{han2021fantasycoref}.
Popular libraries such as bookNLP~\cite{DBLP:conf/lrec/BammanLM20} use the imperfect co-reference models and provide disclaimers that the outcome will be error-prone.
However, in our case, correct co-reference resolution was crucial for capturing the character traits.
We adopted a human-centered approach where a user needs to validate and merge clusters detected in different chapters.
This can be seen as a bottleneck for our system, hindering instantaneous feedback to the user; however, we believe it opens up several new research directions.

The first is to open up the annotation interface (Figure~\ref{fig:annotate_interface}) to users, allowing writers and scholars to annotate the stories themselves.
In our study, we showed the participants a short demo of the Annotation Interface.
Their feedback suggests that naming and merging the mentions is not a complicated process and could be an excellent introspective exercise for writers and scholars.
Another possible solution is to crowdsource the annotation task.
The overall task can be divided into a micro-task of validating a single cluster in a chapter, a common approach in crowdsourcing~\cite{DBLP:conf/uist/KitturSKK11, kittur2013future}.
The micro-tasks can then be aggregated to obtain a fully annotated story.
Such crowd support can be instantaneous, as previously shown by authoring tools with crowd support~\cite{DBLP:conf/uist/BernsteinLMHAKCP10}. 

\textbf{Interactive Error Correction.} 
Most of our models are state-of-the-art and provide high accuracy; however they are not 100\% accurate. Our study suggests writers and scholars will likely be able to identify such errors with their domain knowledge. 
This opens up the opportunity to study how participants can interactively fix errors while using our tool.
We also anticipate that such errors will diminish as the models improve.

\subsection{\revise{Future Work}}

\textbf{Scaling to Multiple Documents.}
\toolname{} currently supports the analysis of a single document.
We believe extending the analysis to multiple documents will lead to several novel applications.
First, writers in our study requested version comparison so that they can easily see the impact of the edits.
Support for such iterative revision has previously been shown to be useful~\cite{du-etal-2022-read}.
Another potential direction is supporting the analysis of a text corpus (i.e., distant reading).
For example, we can answer questions such as \textit{what is the representation of females in stories written during 1900?}
We believe the currently supported trait indicators will provide a more comprehensive view than existing bias detection methods which often focus on one or two traits only~\cite{hoyle-etal-2019-unsupervised}.

\textbf{Extending Supported Trait Indicators.} 
One limitation of \toolname{} is that the supported trait indicators may not be exhaustive. However, our interface is modular and can easily be extended to any number of traits. For example, we can enhance speech analysis by integrating indirect speeches~\cite{papay-pado-2019-quotation}, a third-person narration of discourse, for the characters. Similarly, we can integrate social ties between characters (e.g., parents, brothers) as a new indicator~\cite{elson-etal-2010-extracting}. 

\textbf{Extending Supported Narrative Components.}
While characters are one of the essential components of narrative fiction, it is not the only one.
We believe the design of \toolname{} can be extended to analyze and refine other aspects such as narrative structure, events, time, settings, etc.
For example, one extension could be using a non-linear timeline like StoryCurves~\cite{DBLP:journals/tvcg/KimBISGP18} in \toolname{}.
The main challenge here is detecting the non-linear timeline from an unstructured text.
One possibility is using recurring places (e.g., a castle, or city) to identify the non-linear patterns.
A time detection model~\cite{kim-etal-2020-time} can also be helpful in identifying recurring times.
In the future, we want to empirically evaluate these methods to find ways to capture non-linear structures.

\textbf{Extending to Non-fiction.}
We believe \toolname{} can analyze non-fictional characters too.
While our system is designed based on narrative fiction, it can extract the trait indicators for any recurring entity in the text.
This can be helpful in other domains.
For example, there is a wide interest in analyzing political debates in journalism and digital humanities~\cite{DBLP:conf/ijcai/HaddadanCV19, fivethirtyeight}.
Tools such as \toolname{} can help journalists and scholars find nuance patterns and idiosyncrasies among debate participants and help write analysis reports.
Our future work will focus on understanding the usefulness of \toolname{} in analyzing such non-fictional texts. 

%% file: sections/9conclusion.tex
\section{Conclusion}

We have presented \toolname{}, a web-based visual analytics system for visualizing language indicators of character traits in creative fiction.
The goal is that by so doing, the tool will enable two separate audiences to better understand characterization in a story: the creative writers who produce these stories and the literary scholars and critics who study them.
We began the \toolname{} design process using an initial formative study collecting requirements from professionals representing both of our audiences.
This leads to us formulating several requirements and corresponding design guidelines addressing them.
We then used these findings to build the \toolname{} prototype implementation, including its visualizations, interactions, and NLP components and pipeline.
This yielded a prototype that we used in a summative user study involving 12 representatives from both of our user audiences.
Our findings from this interview-based user study were generally positive and favorable for \toolname{}, suggesting both expected as well as some unexpected uses of the tool for characterization, detecting bias, understanding character arcs, etc.
We thus close this paper with a call to arms for using AI, NLP, and visualization assistance to analyze and understand stories in the future.